\documentclass[preprint,12pt]{elsarticle}

\usepackage{lineno}
\usepackage{amssymb}
\usepackage{url}
\usepackage{longtable,tabularx}
\usepackage{siunitx}
\usepackage{makecell}
\usepackage{multirow}
\usepackage{graphicx}

\usepackage{subcaption}
\usepackage{caption}
\usepackage{array}
\usepackage{amsthm}
\usepackage{amsmath}
\usepackage[most]{tcolorbox}

\newtcolorbox{tabframebox}[1][]{breakable,sharp corners,boxrule=0.5pt,colback=white,halign=center,#1}
\journal{Acta Astronautica}

\begin{document}

\begin{frontmatter}

%% Title, authors and addresses

%% use the tnoteref command within \title for footnotes;
%% use the tnotetext command for theassociated footnote;
%% use the fnref command within \author or \address for footnotes;
%% use the fntext command for theassociated footnote;
%% use the corref command within \author for corresponding author footnotes;
%% use the cortext command for theassociated footnote;
%% use the ead command for the email address,
%% and the form \ead[url] for the home page:
 \title{Machine Learning in Orbit Estimation: a Survey}
%\tnotetext[label1]{Corresponding author}
%% \author{Name\corref{cor1}\fnref{label2}}
%% \ead{email address}
%% \ead[url]{home page}
 \fntext[label2]{PhD Student, NOVA LINCS - Computer Science and Informatics Department}
  \fntext[label3]{Professor, NOVA LINCS - Computer Science and Informatics Department}
\cortext[cor1]{Corresponding author}
%% \affiliation{organization={},
%%             addressline={},
%%             city={},
%%             postcode={},
%%             state={},
%%             country={}}
%% \fntext[label3]{}

%\title{Title of Your Manuscript}

%% use optional labels to link authors explicitly to addresses:
%% \author[label1,label2]{}
%% \affiliation[label1]{organization={},
%%             addressline={},
%%             city={},
%%             postcode={},
%%             state={},
%%             country={}}
%%
%% \affiliation[label2]{organization={},
%%             addressline={},
%%             city={},
%%             postcode={},
%%             state={},
%%             country={}}

\author[inst1]{Francisco Caldas\corref{cor1}\fnref{label2}}
\ead{f.caldas@campus.fct.unl.pt}

\affiliation[inst1]{organization={NOVA School of Science and Technology},%Department and Organization
            addressline={Largo da Torre}, 
            city={Caparica},
            postcode={ 2825-149}, 
            %state={Caparica},
            country={Portugal}}

\author[inst1]{Cláudia Soares\fnref{label3}}
\ead{claudia.soares@fct.unl.pt}
%\author[inst1,inst2]{Author Three}

%\affiliation[inst2]{organization={Department Two},%Department and Organization
%            addressline={Address Two}, 
%            city={City Two},
%            postcode={22222}, 
%            state={State Two},
%            country={Country Two}}

\begin{abstract}
%% Text of abstract
%Since the late ’50s, when the first artificial satellite was launched, the number of resident space objects (RSOs) has steadily increased. It is estimated that around 1 Million objects larger than 1 cm are currently orbiting the Earth, with only 30,000, larger than 10~cm, presently being tracked. To avert a chain reaction of collisions, known as Kessler Syndrome, it is indispensable to accurately track and predict space debris and satellites’ orbits. Current pure physics-based methods have errors in the order of kilometers for 7-day predictions, which is insufficient when considering space debris that have mostly less than 1 meter. Typically, this failure is due to uncertainty around the state of the space object at the beginning of the trajectory, forecasting errors in environmental conditions such as atmospheric drag, and specific unknown characteristics such as the mass or geometry of the Resident Space Object. Leveraging data-driven techniques, namely Machine Learning, the Orbit Prediction accuracy can be enhanced by deriving unmeasured objects’ characteristics and improving non-conservative forces’ effects. % And by utilizing the superior abstraction capacity that Deep Learning models have of modelling highly complex non-linear systems.  In this survey, we provide an overview of the work in applying Machine Learning for Orbit determination, Orbit prediction, and atmospheric density modeling.

Since the late 1950s, when the first artificial satellite was launched, the number of Resident Space Objects has steadily increased. It is estimated that around one million objects larger than one cm are currently orbiting the Earth, with only thirty thousand larger than ten cm being tracked.
To avert a chain reaction of collisions, known as Kessler Syndrome, it is essential to accurately track and predict debris and satellites' orbits. Current approximate physics-based methods have errors in the order of kilometers for seven-day predictions, which is insufficient when considering space debris, typically with less than one meter.
This failure is usually due to uncertainty around the state of the space object at the beginning of the trajectory, forecasting errors in environmental conditions such as atmospheric drag, and unknown characteristics such as the mass or geometry of the space object. Operators can enhance Orbit Prediction accuracy by deriving unmeasured objects' characteristics and improving non-conservative forces' effects by leveraging data-driven techniques, such as Machine Learning.
In this survey, we provide an overview of the work in applying Machine Learning for Orbit Determination, Orbit Prediction, and atmospheric density modeling.

\end{abstract}

%%Graphical abstract
\begin{graphicalabstract}
\includegraphics[width=\columnwidth]{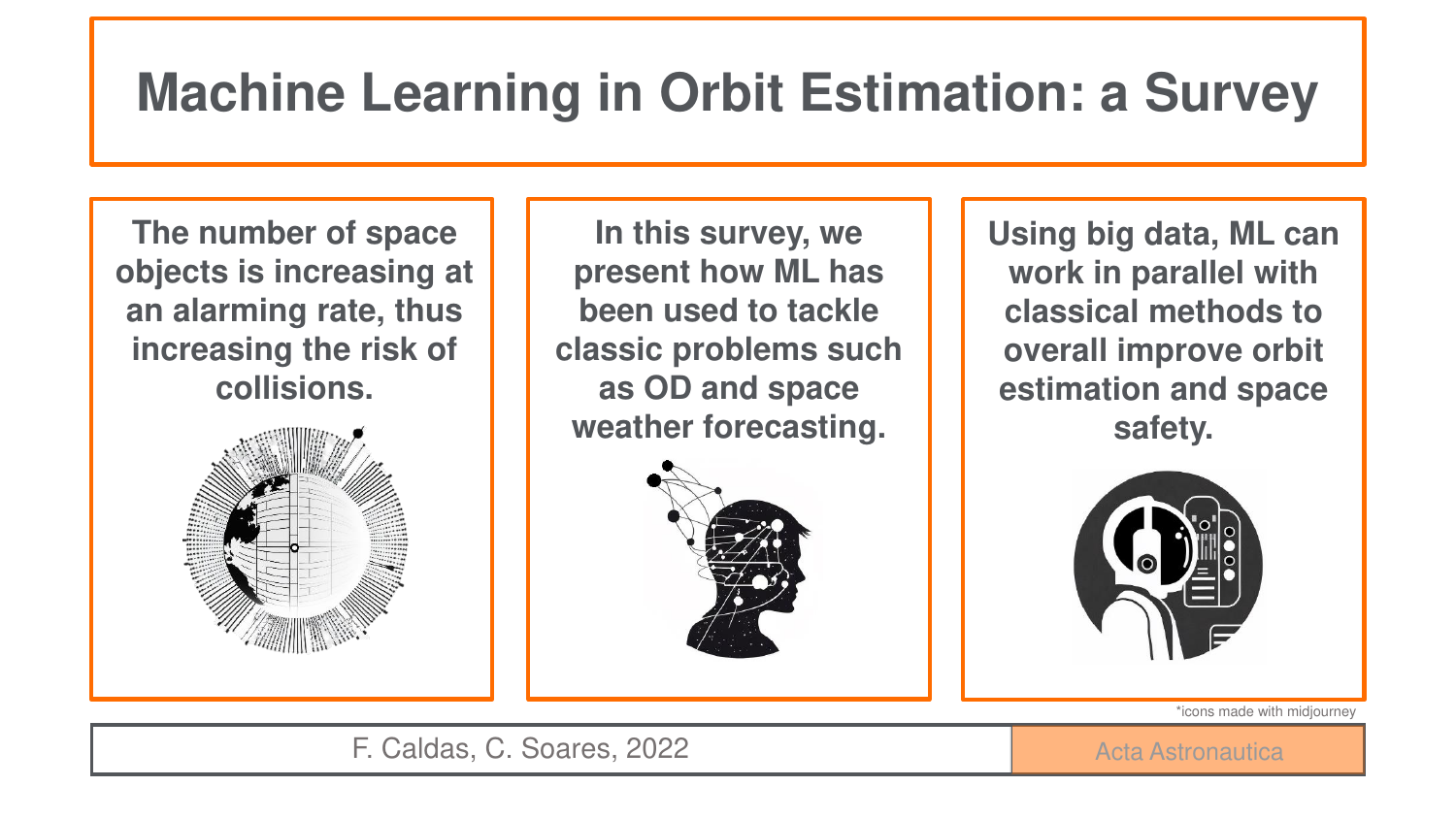}
\end{graphicalabstract}

%%Research highlights
\begin{highlights}
%\item Summarize the physical forces and equations that define the models currently in use.
\item Survey the introduction of new statistical techniques for Orbital Determination and Orbital Prediction.
\item Understand the properties of empirical atmospheric density models and how Deep Learning is applied in this context.
\item Systematically study trends in Orbital Estimation using Machine Learning-based enhancements.
\item Define the limitations, and potential applications of Machine Learning approaches to Orbit Determination.

\end{highlights}

\begin{keyword}
%% keywords here, in the form: keyword \sep keyword
Orbital Mechanics \sep Machine Learning \sep Deep Learning \sep Low-Earth Orbit \sep Satellites \sep Atmospheric Density Models \sep Orbit Determination \sep Orbit Prediction
%% PACS codes here, in the form: \PACS code \sep code
\PACS 01.30.Rr \sep 07.05.Mh
%% MSC codes here, in the form: \MSC code \sep code
%% or \MSC[2008] code \sep code (2000 is the default)
\MSC[2008] 68-02 \sep 68T37
\end{keyword}

\end{frontmatter}

%% \linenumbers

%% main text

%\linenumbers

\begin{tcolorbox}[arc=0pt,colback=white]
\textbf{List of Acronyms}
{\renewcommand\arraystretch{0.9}
\noindent\begin{longtable}{p{3cm}l}
ANN & Artificial Neural Network \\
CNN & Convolutional Neural Network \\
ECI & Earth-Centered Inertial \\
EKF & Extended Kalman Filter \\
EUV & Extreme Ultraviolet light \\
FNN & Feed-Forward Neural Network \\
FPKE & Fokker-Planck-Kolmogorov Equation \\
GBDT & Gradient Boosting Decision Tree \\
GMM & Gaussian Mixture Model \\
GP & Gaussian Process \\
GPS & Global Positioning System\\
GS & Ground Station \\
GVM & Gauss-Von Mises \\
HAMR & High Area-to-mass Ratio \\
LEO & Low Earh Orbit \\
LFM & Latent Force Model \\
LSTM & Long-Short Term Memory \\
ML & Machine Learning \\
OD & Orbit Determination \\
OP & Orbit Prediction \\
PF & Particle Filter \\
PINN & Physics-Informed Neural Network \\
PoC & Probability of Collision \\
RF & Radio Frequency \\
RNN & Reccurent Neural Networks \\
RSO & Resident Space Object \\
SDE & Stochastic Differential Equation \\
SGP4 & Simplified General Perturbations model 4 \\
SVM & Support Vector Machine \\
TDNN & Time-Delay Neural Networks \\
TLE & Two-Line Element \\
UKF & Unscented Kalman Filter \\
\end{longtable}}
\end{tcolorbox}
\section{Introduction}

%\begin{linenumbers}
It is estimated that more than 36,000 objects larger than 10~centimeters and millions of smaller pieces exist in Earth's orbit~\cite{esa_2021}. To safeguard active spacecraft, it is necessary to accurately determine where each Resident Space Object (RSO) is and where it will be at all times. To create such a complex body of knowledge, the more broad problem of orbit estimation is divided into sub-problems, each largely complex but with more specific goals. In this review, we observe three main sub-problems: Orbit Determination, Orbit Prediction, and Thermospheric Mass Density. 

\begin{itemize}

\item \textbf{Orbit Determination (OD):} the OD is the determination of the orbit of the object based on observations~\cite{Vallado2001,curtis2014}. The Extended Kalman Filter (EKF)~\cite{Smith1962,Julier1997,Einick} is the de facto standard for orbit determination in real-world scenarios~\cite{tapley2004}. The accuracy of this process depends on the number of sequential observations used to determine the orbit, and the type of observation, e.g., laser ranging and GPS tracking, which is far more precise than optical observations. The output of this method is commonly a state vector of the orbit of the object, usually represented through a vector consisting of the object's estimated position and velocity and a covariance matrix reflecting the uncertainty under the Gaussian assumption. Currently, this process is limited by the assumptions of the EKF, a lack of knowledge of RSO's shape and attitude, and dynamic model simplifications, which we will further examine in Section~\ref{chap:od}.

\bigskip

\item \textbf{Orbit Prediction (OP):} Orbit Prediction is the process of predicting the future position and associated uncertainty of any given RSO. Two method families exist for OP: one pursuing analytical solutions~\cite{Montenbruck2005,bate1971fundamentals}, with the other exploiting numerical approximations~\cite{UPHOFF1972}. Numerical methods are time-consuming but precise, while analytical methodologies are more straightforward and faster. To be tractable, these algorithms hold simplifying assumptions that hinder accuracy~\cite{Vallado2001}. Each method is limited by OD in that an orbital state with high uncertainty will necessarily evolve to have an inaccurate Orbit Prediction. When used in a Collision Avoidance scheme, this process propagates the state of the RSO until the time of closest approach (TCA) to any actively monitored satellite. The current limitations in Orbit Prediction are the Gaussian assumption --- which does not hold against the true distribution~\cite{Poore2016} --- the simplified modeling of perturbation forces, the unknown information regarding RSO characteristics, and the uncertainty over space weather forecast data.

\bigskip

\item \textbf{Thermospheric Density Mass Models:} A set of force models determine the acceleration of any RSO. Earth's gravity potential is the most meaningful, but solar and lunar gravitational attraction and Earth/ocean tides affect the RSO~\cite{Montenbruck2005}. Of the non-conservative forces, air drag is the most relevant for objects in LEO. Being applied in the opposite direction of the RSO's velocity, this force is the largest source of uncertainty for most RSOs in LEO~\cite{Storz2005}, and correctly determining atmospheric density is vital to compute satellite drag.
All state-of-the-art models currently used for density estimation are empirical (data-driven) models that have been consistently updated since the last century. The lack of predictive capability and uncertainty estimation are the two main drawbacks of these methods.

\end{itemize}

\begin{table}[hb!]
  \caption{\label{tab:table1} Outlining of input and output for each task}
    \begin{tabular}{lll}
      \hline
      Task & Input & Output \\ 
      \hline
      Orbit Determination & Ground Station (GS) obs. & {\makecell{orbital states \\ at time $t_0$ }} \\
      Orbit Propagation & orbital states & {\makecell{orbital states \\ at time $t_1$ }}\\
      Thermospheric Density & Space weather $(F_{10.7},ap, ..) $&\makecell{ local density \\ and temperature}\\ \hline
    \end{tabular}
    %\vspace{.5em}
    %\begin{tablenotes}\small
    %\item[*]  and associated uncertainties. 
    %\end{tablenotes}
\end{table}
%%%
In Table \ref{tab:table1}, we have a summary of the input/output for each task. In the following sections, we will thoroughly review published work using machine learning techniques to help solve each of the problems mentioned previously. Each section will start with a brief description of the task, followed by a state-of-the-art review of the current classical methods. The machine learning body of work will be presented in semi-chronological order, from the initial use of historical data to improve physical models to the most recent developments in the area that use highly complex models. The following section will briefly define the physical forces that determine the orbit of a satellite or space debris.
%\begin{tcolorbox}[arc=0pt,colback=white]
%\textbf{Nomenclature}
%{\renewcommand\arraystretch{0.9}
%\noindent\begin{longtable}{@{}l @{\quad=\quad} l@{}}
%$A_k$ & local linearization of the dynamics function $f$\\
%$ap$ & geomagnetic index \\
%$a$  & cross-sectional area \\
%$\textbf{a}_{pert}$ &    perturbing acceleration \\
%$BC$ & ballistic coefficient \\
%$c_D$& drag coefficient \\
%$F_{10.7}$ &  solar radio flux at $10.7$ cm \\
%$F_{30}$ & $30~cm$ solar radio flux \\
%$f$   & system dynamics vector \\
%$H_k$ & local linearization of the measurement function $h$\\
%$h$  &  measurement function\\
%$k$  & time index \\
%$K_k$ & Kalman gain at time $k$ \\
%$m$ & space object's mass \\
%$Q_k$ & process noise covariance at time $k$ \\
%$R_k$ & measurement noise covariance at time $k$ \\
%$r$ & position coordinates \\
%$\dot{r}$ & velocity coordinates\\
%$\ddot{r}$ & acceleration coordinates\\
%$t$ &  time \\
%$t_0$ & initial time \\
%$v_{rel}$ & relative velocity to the rotating atmosphere \\
%$v_k$ & measurement Gaussian white noise  \\
%$x$ & state vector \\
%$y_k$ & observations \\
%$w(t)$ & process Gaussian white noise \\
%$\mu$ & mean of a Gaussian Distribution \\
%$\mu_\oplus$ & gravitational constant times mass \\
%$\Sigma$ & variance of a Gaussian Distribution \\
%$\Phi$ & State Transition Matrix (STM) \\
%$\phi$ & flow solution\\
%$\rho$ & atmospheric density mass\\
%\end{longtable}}
%\end{tcolorbox}

%\end{linenumbers}
\section{Forces Model}

%\end{linenumbers}
The equation of motion of a space object in a Cartesian ECI (Earth-Centered Inertial) coordinate system~\cite{Horwood2011,Vallado2001} can be written in the form
\begin{equation}
\label{sample:2}
f(\boldsymbol{x},t) = \ddot{\boldsymbol{r}} = - \frac{\mu_\oplus}{r^3}\boldsymbol{r} + \boldsymbol{a}_{pert}(\boldsymbol{r},\dot{\boldsymbol{r}},t) ,
\end{equation}
where $r = ||\boldsymbol{r}||$ is the norm of the position vector, $[\boldsymbol{r} \quad \dot{\boldsymbol{r}}]^T= \boldsymbol{x} $, and $\mu_\oplus$ is the gravitational constant multiplied by the masses of the Earth and the RSO, with the perturbing forces being
\begin{equation}
    \boldsymbol{a}_{pert}(\boldsymbol{r},\dot{\boldsymbol{r}},t) = \boldsymbol{a}_{NS} + \boldsymbol{a}_{S/M} + \boldsymbol{a}_{drag} + \boldsymbol{a}_{SRP} + \boldsymbol{a}_{tides}  + \boldsymbol{a}_{others} .
\end{equation}
For a given initial condition $\boldsymbol{x}(t_0) = \boldsymbol{x}_0$, the position of a satellite at any point in time can be implicitly written as the solution of the equation flow $\phi$:
\begin{equation}
\boldsymbol{x}(t) = \phi(t;\boldsymbol{x}_0,t_0).
\label{equation:flow}
\end{equation}
The perturbation of Earth's gravity potential, $\boldsymbol{a}_{NS}$, is due to the fact that Earth is not a perfect sphere. This force is modeled using spherical harmonic potential equations, with varying levels of precision depending on the degree. The most recently published model, the Earth Gravity Model (EGM2012), has coefficients up to degree 2190 and is accurate to $0.1$m~\cite{pavlis2012}. 
The $n$-body perturbations, $\boldsymbol{a}_{S/M}$, affect the trajectories of RSOs in orbit, with the Solar and Lunar perturbations being the two most relevant forces on the space object's body~\cite{Vallado2006}. Other planets in the solar system, such as Jupiter, have minimal impact on satellites.
The atmospheric drag, $\boldsymbol{a}_{drag}$, especially in the Thermospheric layer ($85/125 - 600/1000$ km), has non-negligible effects, and it is responsible for the decay of orbits in Low Earth Orbit (LEO).
Using the canon-ball model, the aerodynamic drag can be represented by~\cite{Doornbos2012}:
\begin{equation}
\label{equation:drag}
    \boldsymbol{a}_{drag} = -\frac{1}{2} \rho \frac{ A_{ref}c_{D} }{m} v_{rel}^2\boldsymbol{\hat{v}_r}
\end{equation}
where $\rho$ is the local atmospheric mass density, $c_D$ is the drag coefficient, $A_{ref}$ is the cross-sectional area, $m$ the mass of the space object, $v_{rel}$ the magnitude of the relative velocity of the object with respect to the rotating atmosphere and $\boldsymbol{\hat{v}_r}$ the unit vector of the relative velocity. For most satellites, the mass and cross-sectional areas are known in advance and are assumed to be constant for the lifespan of the satellite, despite slight changes that might occur due to, for example, fuel consumption. However, this is not the case for most RSOs, especially space debris originating from breakups. Taking that into account, 
%Variables such as mass and cross-sectional area are assumed to be constant and known, and while this is true for most satellites, it is not valid for space debris. The drag coefficient is the most challenging variable to determine~\cite{gaposchkin1994calculation,Vallado2014a} being defined by a combination of molecular content, reflection, attitude, and altitude. Because these object-specific parameters are unknown for most RSO in the catalogue,
it is common to define a combined parameter, the Ballistic Coefficient ($BC = \frac{m}{A_{ref} c_{D}})$, which integrates mass, area and drag coefficient to be calculated as an extra feature during Orbit Determination. The local mass density $\rho$ is the variable obtained from the Thermospheric Density Mass models, which is particularly challenging to model, as we will see in Section \ref{sec:ther}.

The Solar Radiation Pressure, $\boldsymbol{a}_{SRP}$, is the force that arises from the absorption and reflection of light from the Sun. It depends on the cross-sectional area and mass, but unlike atmospheric drag, it does not depend on altitude, but on the Sun-space object distance. It is the dominant non-gravitational perturbation on higher altitudes (MEO and GEO) and for High Area to Mass Ratio (HAMR) objects~\cite{Poore2016}. Of particular importance to the representation of this force is the determination of the entry and exit times of a given RSO from the shadow of the Earth, along with the determination of the force orientation, as the object-Sun vector is constantly changing and, with it, the cross-sectional area.

Tidal effects, $\boldsymbol{a}_{tides}$, are also taken into account in orbital determination, with the small periodic deformations of the solid body of the Earth referred to as solid Earth tides, which, alongside ocean tides, impact Earth's gravity potential~\cite{Montenbruck2005}. Other perturbations include General Relativity theory adjustments, geomagnetic pulls, Earth's albedo, and atomic clock corrections.

\begin{figure}[!b]
    \centering
    \includegraphics[width=0.9\columnwidth]{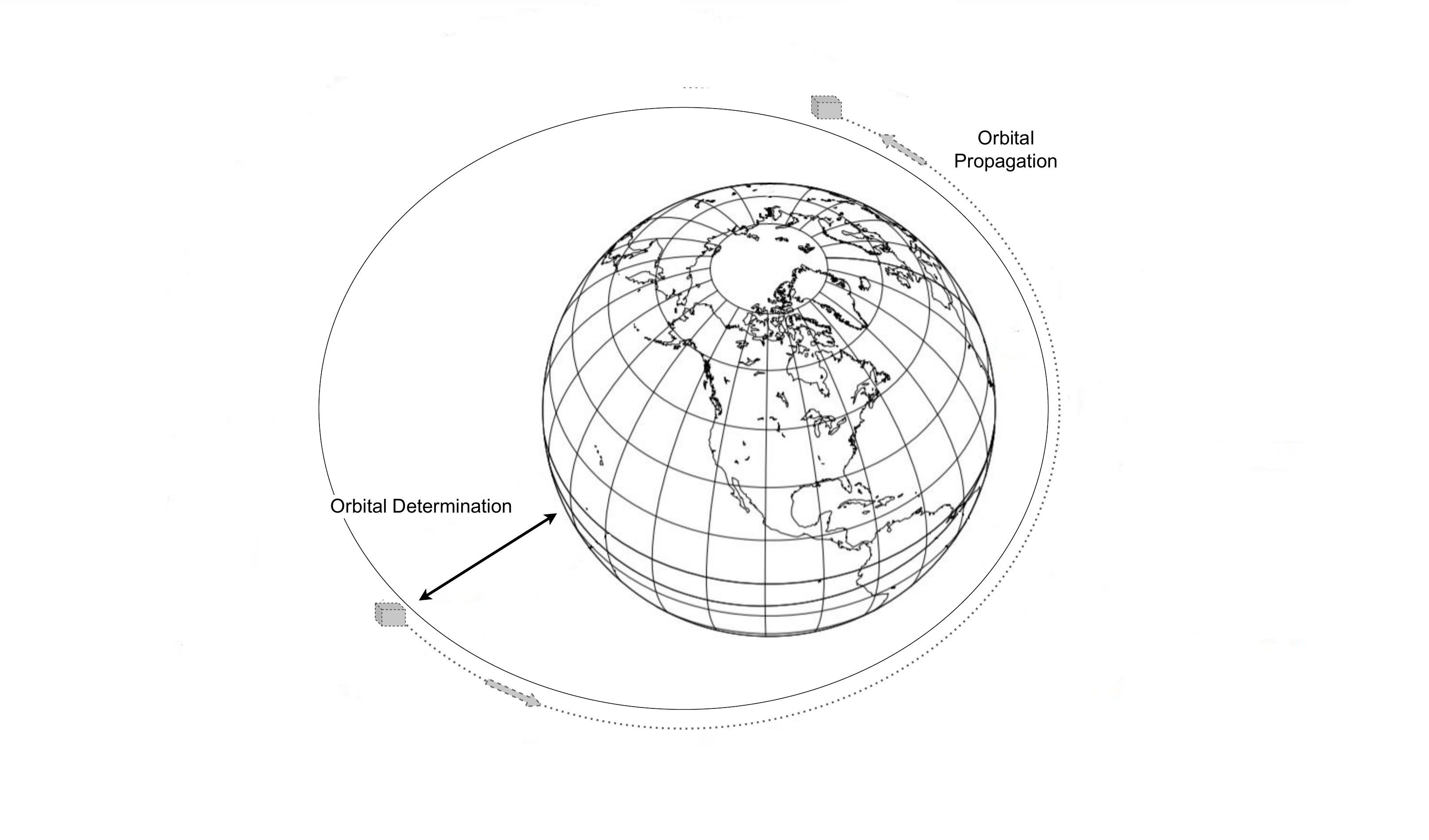}
    \caption{The two main stages of orbital estimation necessary for collision avoidance and debris and satellite tracking. The information obtained from OD is used as input for Orbit Prediction (OP). Figure adapted from~\cite{Vallado2001} and~\cite{Luo2017}.}
    \label{fig:odvsop}
\end{figure}

\section{Orbit Determination}
\label{chap:od}

This is the process of using observations to obtain a state vector $\boldsymbol{x} \in \mathbb{R}^p$ of a given space object~\cite{Milani2009}. In this survey, space objects will be all objects, either satellites or space debris, currently orbiting the Earth. A state vector of $p=6$ is sufficient to represent the velocity and position of the satellite at a given time. Still, other variables, such as the Ballistic Coefficient (BC) or clock bias, can also be calculated as part of the state vector~\cite{Mortlock2021}. This state vector can also be represented by the traditional Keplerian elements $(e,a, i, \Omega, \omega, \theta) $ or by equinoctial elements $(a, h, k, p, q, \ell)$. The types of observations used for this purpose are radio, radar, laser, or image, with observations usually divided between cooperative or non-cooperative observations~\cite{Luo2017}. As the name suggests, cooperative observation is when the object in question, e.g., a satellite, has a built-in capability to respond to tracking. This gives the Ground Station (GS) the ability to know with detail the position and ID of the object. In this case, we have a significant amount of observations across the period in which the GS, or GS's, communicate with the satellite. The non-cooperative observations are necessary to track defunct satellites or space debris. In this case, identifying the satellite is a necessary part of the process, and the number of observations, usually radar, is relatively small for each space object.

The state-of-the-art approach for this problem is the Extended Kalman Filter (EKF)~\cite{Vallado2001}. This method approximates the Kalman Filter to non-linear systems, such as the equations of motion of a space object~\cite{tapley2004}. Its limitations include using only the first-order Taylor series expansion to approximate the system to a linear system and the assumption that the measurement and process noise is Gaussian~\cite{Krener2007}. Despite that, the applicability of the filter for sequential data, alongside providing a probability distribution of the estimated state, proves very useful and justifies the current use. Because of the relevance of the EKF method to orbit determination, we will briefly explain the algorithm.

\begin{figure}[!b]
    \centering
    \includegraphics[scale=0.6]{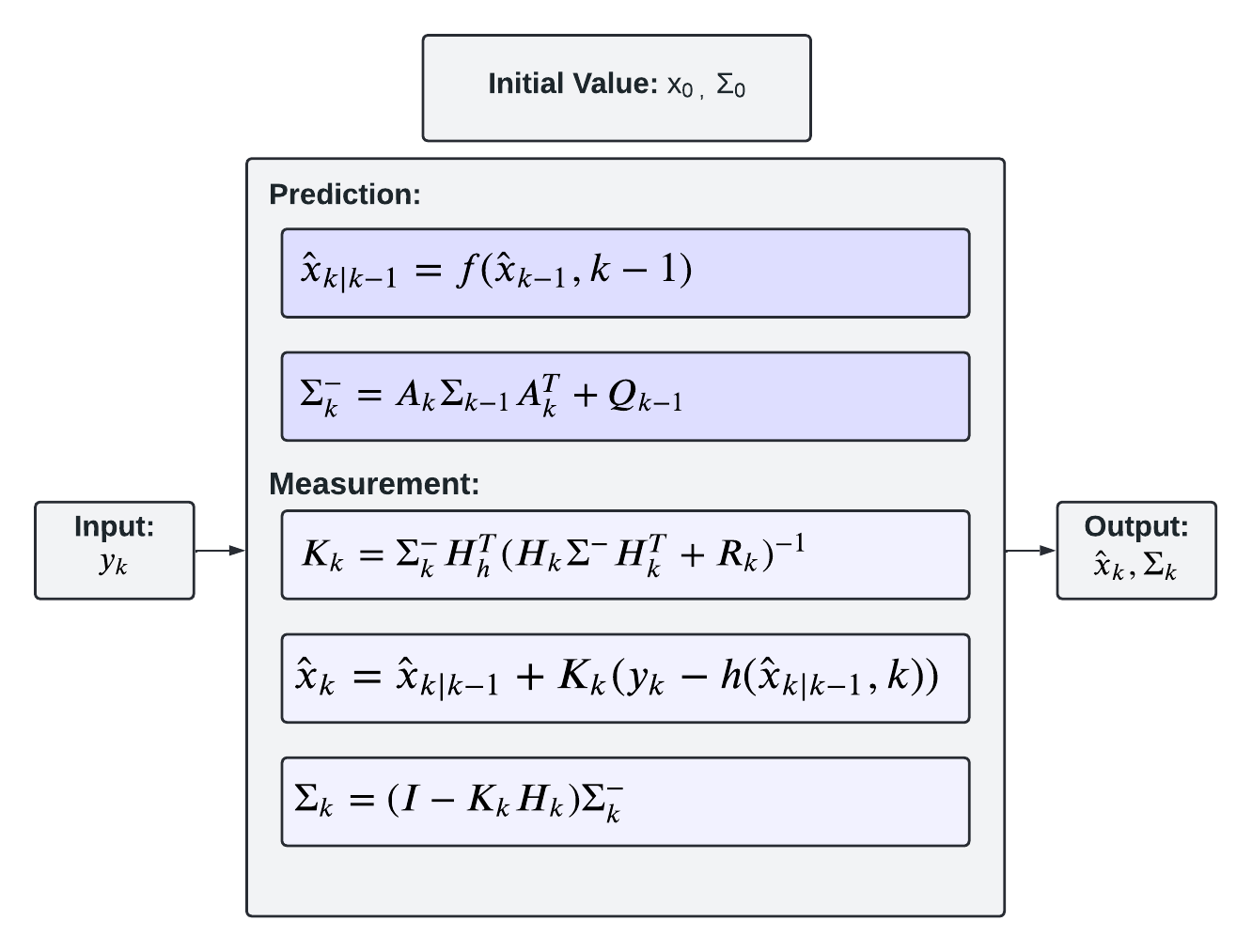}
    \caption{General Kalman Filter Diagram. $A_k$ is the state transition matrix at time $k$, $K_k$ is the Kalman Gain at each observational moment, $H_k$ is the linear measurement mode at time step $k$, and the output is the estimated mean ($\hat{x}_k$) and covariance ($\Sigma_k$) at time step $k$. $\hat{x}_{k|k-1}$ and $\Sigma^-_k$ are internal components representing the \textit{a priori} prediction before the observation $y_k$ is introduced.}
    \label{fig:kalman}
\end{figure}

The intended goal of this method is to combine knowledge of the dynamical system with observations. Therefore the dynamical and observation equations can be represented as~\cite{Krener2007,kalman1960}:
\begin{equation}
    \begin{split}
  \dot{\boldsymbol{x}}(t) &= f(\boldsymbol{x},t) + w(t) \\
  y_k &= h(x_k) + v_k
  \label{eq:jf}
  \end{split}
\end{equation}
where $w(t) \sim N(0,Q(t))$ is the process noise, $y_k$ is the measurement, $h(x_k)$ the measurement function and $v_k$ the measurement noise following a zero-mean Gaussian with covariance $R_k$, $v_k \sim N(0,R_k)$. Through this general formula, we can observe that the formula combines discrete observations with continuous dynamical equations. When $f(\boldsymbol{x},t)$ and $h(x_k)$ are linear functions, and the initial distribution is Gaussian, with $w(t)$ and $v_k$ expressing white noise (uncorrelated Gaussian distributions with zero-mean), the Kalman Filter is the optimal filter~\cite{Ribeiro2004}. A known and much-studied filter in the automatic control and time-series research fields, many books~\cite{Gelb1974,Kim2011,Bang18} present in-depth explanations of the Kalman filter, which can be summarized as a two-step algorithm (predict and filter) as illustrated in Figure~\ref{fig:kalman}.

The Extended Kalman Filter is applied when the linearity assumption does not hold, particularly the linearity of $f(\boldsymbol{x},t)$ and/or $h(x_k)$. To do so, and before the KF algorithm in Figure \ref{fig:kalman} ought to be used, both equations are linearized using the first order of Taylor's approximation at each observational moment $k$:
\begin{equation}
    \begin{split}
  f(x_k) &\approx f_k(\hat{x}_{t|k})+ \frac{\partial f(x_k,t)}{\partial x_k}\Bigr{|}_{x_k = \hat{x}_{t|k}} + ... \\
  h(x_k) &\approx h_k(\hat{x}_{t|k})+ \frac{\partial h(x_k)}{\partial x_k}\Bigr{|}_{x_k = \hat{x}_{t|k}} + ... \\
  \end{split}
\end{equation}
%with the KF algorithm presented in \ref{fig:kalman} . 
Note in Figure~\ref{fig:kalman} that the prediction step for the \textit{a priori} mean ($\hat{x}_{k|k-1}$) does not need a linear function. It is the covariance matrix and the Kalman Gain matrix that require the simplification of the dynamical and measurement equations. This is precisely the intuition behind improvements over the base method such as the Unscented Kalman Filter (UKF)~\cite{Julier1997}, the Second Order Kalman Filter~\cite{Einick}, and others methods referred to in Lee~\cite{Lee2005}. These methods more closely represent the nonlinearity of the problem but are, nonetheless, Gaussian or linear approximations. Sequential Monte Carlo methods such as the Unscented Particle Filter~\cite{payne2004} or the Divided Difference Particle Filter~\cite{Lee2005} are also popular and shown to be superior to the EKF by~\citet{Arulampalam2002} and~\citet{Ning2012}, with the disadvantage of high computational costs~\cite{Ning2012}. The UKF, in particular, sits on the border between Monte Carlo and linear methods, with deterministically chosen sigma-points ($\mathcal{X}$)~\cite{Julier1997}, made to represent the Gaussian distribution, being independently propagated across time through $x_{k+1} = \phi(t_{k+1};x_k,t_k)$ and then used to derive the mean and covariance at the $k+1$ time step. 

More recently, the problem's non-linearity has been tackled through machine learning and data-driven techniques. Hartikainen et al.~\cite{Hartikainen2012} used Latent Force Models (LFMs) to combine the dynamic system with a more freely defined measurement component, implementing Gaussian Processes (GP) to model measurement noise and using a machine learning approach to define the covariance structure of the Gaussian Process~\cite{alvarez2019,sarkka2019}. 
%A different method was introduced in~\cite{Hartikainen2012} using Latent Force Models (LFMs) to combine the physical models principles with non-parametric data-driven components.
This method improves on the Kalman Filter by removing the uncorrelated white noise assumption and replacing it with a time-varying GP. 
A different method uses Physics Informed Neural Networks (PINN) to incorporate physical knowledge into the optimization process of the Neural Network~\cite{pinnsProposed}. This approach has been applied to the circular restricted three-body problem \cite{ghilardi} and data from cislunar objects \cite{scorsoglio2023physics}.
In~\citet{Sharma2015}, it is presented a general approach to OD using distribution regression~\cite{szabo2016learning}. It is shown that under a known and continuous dynamic system, with known spacecraft characteristics and observable orbital parameters, there is a possible mapping from the probability distribution of the observed samples to the orbital parameters. This method can be used either to estimate the posterior distribution of the orbital parameters based on full batch observations or to identify the ID of the spacecraft associated with each sample. In~\cite{Sharma2018}, Sharma extends this work and presents a method for collaborative and non-collaborative orbit classification by estimating the probability of each sample to represent the distribution associated with its ID. This technique is particularly useful to 
%Using an optimization function over a set of meta-distributions, the authors are able to %estimate a kernel embedding associated with the observed samples. Furthermore, in is Thesis,  %extends this work and shows that the general method can be used for the classification of space objects. Using transfer learning, the estimated regression distributions can be used
to identify satellites with identical Radio Frequency (RF) transmissions or very close orbits that were not extended to more than two spacecraft. The distribution regression approach was also followed in~\citet{Jiang2021}, extending the process to observations from more than one Ground Station (GS).

Different authors~\cite{Terejanu2008,DeMars2013,Horwood2011} independently presented an improvement over the current methods for OD by using Gaussian Mixture Models (GMMs) to model the estimated distribution. A Gaussian Mixture model is a mixture distribution created by combining multiple Gaussian distributions, as represented in \eqref{eq:gmm}:
\begin{equation}
    p(\theta) = \sum_{i=1}^k w_i\mathcal{N}(\mu_i,\Sigma_i)
    \label{eq:gmm}
\end{equation}
with $\sum_{i=1}^k w_i =1$. A GMM is known to approximate any other distribution given a sufficient number $k$~\cite{Alspach1972,GoodBengCour16}, and therefore it is used to approximate the real distribution of the state during the OD period. This allows for a more accurate characterization of the distribution of $X$, being $X$ the state at any given time during OD, while also being particularly useful for the step of Orbit Propagation (OP), as all current methods are capable of propagating a Gaussian distribution across time, and are trivially scalable to a finite number of independent Gaussian distributions. This work was extended by~\citet{Vishwajeet2014} and~\citet{Terejanu2011} by adapting the GMM weights across time, and further enhanced by Vittaldev~\cite{Vittaldev2015,Vittaldev2016} who presented a univariate splitting library along arbitrary directions that are used to generate a Multidirectional Gaussian Mixture Model. Horwood et al.~\cite{Horwood2012} also proposed a new distribution to improve OD --- the Gauss von Mises (GVM) distribution --- using this distribution in tandem with a particular set of orbital elements, the $J_2$ equinoctial orbital elements to reduce the non-linearity of the process. The GVM is the counterpart distribution of the Gaussian distribution for circular support.
%a circular Gaussian approximation, meaning that it approximates the wrapped normal distribution in the unit circle.
In a series of publications, Horwood et al.~\cite{Horwood2014,horwood2014a,Poore2016} compare their approach with the UKF, EKF, and GMM methods using a Particle Filter (PF) as the ground truth model. Both the GMM ($k=49$) and the GVM are shown to faithfully represent the orbit distribution eight times longer than the UKF and the EKF.

%\cite{Lee2018} introduces a new method for Initial Orbit Determination using automatic machine learning. Using an initial set of the satellite's range rate, and satellite pass duration, this machine learning methodology is used to predict the range rate for a future satellite pass. talvez meter o kozhaya e o mortlock aqui

\begin{figure}
   \centering
   \begin{subfigure}{0.48\textwidth}
         \centering
         \includegraphics[width =\linewidth]{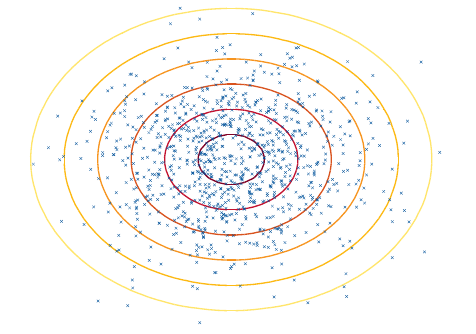}
    \caption{Initial Gaussian representation of the uncertainty}
         \label{fig:2}
     \end{subfigure}
     \hfill
     \begin{subfigure}{0.48\textwidth}
         \centering
         \includegraphics[width=\textwidth]{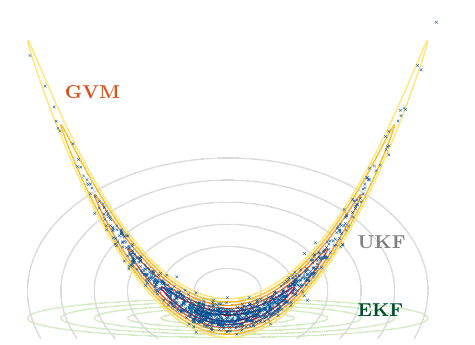}
         \caption{Propagated uncertainty Using the EKF, UKF, GVM filter and particle filter, after 8 orbital periods}
         
         \label{fig:3}
     \end{subfigure}
    \caption{Uncertainty representation comparison between different filters for Orbit Determination. The blue marks are the ground truth representation using a PF. Image in \citet{Poore2016}.}
    \label{22}
\end{figure}

\section{Orbit Prediction}

Orbit Prediction can usually be seen as propagating a known orbital state to a future orbital state. The current physics-based methodology is to solve the differential equations of motion analytically or numerically to predict the mean state at a future time. The numerical methods are time-consuming, and the sheer number of space debris objects orbiting the Earth makes them only usable for academic or non-real-time applications~\cite{UPHOFF1972,bate1971fundamentals}. In general, the non-linear stochastic dynamic system can be defined using the Itô stochastic differential equation (SDE)~\cite{Risken_1989,curtis2014}:
\begin{equation}
\dot{\boldsymbol{x}} = f(\boldsymbol{x},t)dt + G(t)d\beta(t),
\label{eq:sde}
\end{equation}
where $f(\boldsymbol{x},t)$ is the deterministic part of the system, as seen in Equation \eqref{sample:2}, $G(t)$ is the diffusion of the process, and $\beta(t)$ is a Brownian motion, also known as a Wiener process, with zero mean and covariance $Q$. Given an initial probability distribution of $\boldsymbol{x}$, $p(\boldsymbol{x}_0,t_0)$, the pdf at a future time $p(\boldsymbol{x},t)$ is governed by the Fokker-Planck-Kolmogorov equation (FPKE)~\cite{jazwinski2007stochastic}:
\begin{equation}
    \frac{dp}{dt} = -\nabla^T_{\boldsymbol{x}}(pf) + \frac{1}{2}\text{tr}[\nabla_{\boldsymbol{x}}\nabla_{\boldsymbol{x}}^T(pG(t)Q(t)G(t)^T)],
    \label{eq:sdo}
\end{equation}
where $\nabla_{\boldsymbol{x}}$ is the gradient in $\boldsymbol{x}$ viewed as a column operator. Solving this equation would provide a complete description of the position and uncertainty of the orbit at any time $t>0$. However, there is no analytical solution to most FPKEs, with orbital mechanics being particularly hard to solve due to the nonlinearity of the perturbed dynamics in a state space with at least six dimensions. The Orbit Prediction numerical approximations of this solution usually entail a set of assumptions to allow the problem to be solvable. In this review, before jumping to the advances in the field, we are going to present the more common approach for Orbit Prediction, where the output from the EKF, a Gaussian distribution with mean $\mu(t_0) = \mu_0$ and covariance $\Sigma_0$ is propagated according to a first-order system of SDEs. Using the SDE in~\eqref{eq:sde}, assuming the noiseless case $(Q(t)=0)$ and the flow solution \eqref{equation:flow}, the mean and covariance at time $t$ is:
\begin{equation}
    \begin{split}
        \mu(t) &= \phi(t;\mu_0,t_0) \\
        \Sigma(t) &= \Phi(t;\mu_0,t_0)\Sigma_0\Phi(t;\mu_0,t_0)^T ,
    \end{split}
\end{equation}
where the flow solution $\phi(t;\mu_0,t_0)$ and the State Transition Matrix (STT) \\ $\Phi(t;\mu_0,t_0)$ are solved by:
\begin{equation}
    \begin{split}
        \phi'(t;\mu_0,t_0) &= f(\mu_0,t) \\
        \Phi'(t;u_0,t_0) &= F(\phi(t;\mu_0,t_0),t) \Phi(t;u_0,t_0) ,
    \end{split}
    \label{eq:ode}
\end{equation}
where $\Phi(t_0;u_0,t_0) = I$ and, in the case of the EKF, the linearly propagated covariance is obtained with $F(\phi(t;\mu_0,t_0),t)$ as the partial derivative of the function $f(\boldsymbol{x},t)$ in Equation \eqref{sample:2}:

\begin{equation}
    F(\boldsymbol{x},t) = \frac{\partial f(\epsilon,t)}{\partial \epsilon} \Bigr|_{\epsilon=\boldsymbol{x}}.
\end{equation}

To solve the equations in \eqref{eq:ode} there are many numerical methods, with some numerical integration methods specialized for orbit propagation, as seen in~\citet{Montenbruck2005} and~\citet{Jones2012}. The most relevant numerical integrators used for orbit propagation found in the literature include explicit Runge-Kutta methods such as Dormand-Prince 8(7) (DP8) or Runge-Kutta-Nystrom 12(10) (RKN12) and predictor-corrector methods, namely Adams-Bashforth-Moulton (ABM) and Gauss-Jackson (GJ)~\cite{Poore2016}. In particular, implicit Runge-Kutta methods such as the ones developed by Bradley et al.~\cite{Bradley2012} (GL-IRK) and Bai and Junkins~\cite{Bai2011} (GC-IRK), are interesting algorithms for practical applications as they allow for parallelization, are A-stable~\cite{Butcher} at all orders and are more accurate than DP8~\cite{Aristoff2014}. To increase the stability and precision of these numerical methods, regularized orbital formulations, such as the Kustaanheimo-Stiefel (KS)~\cite{Sharma1988,Kustaanheimo} transformation, the Sperling-Bürdet regularization~\cite{sperling1961computation,burdet1967regularization} and the DROMO~\cite{bau2014,Urrutxua_2015} formulation were developed so as to reduce the impact of the non-linearity and singularities of the problem.
Analytical methods are more common, and of those, the widely used method is the SGP4 (Simplified General Perturbations Model 4)~\cite{Zwiep2009}. Made available to the public in 2006~\cite{Vallado2006}, it has since become a staple for Orbit Prediction, despite the lack of accuracy for long-time predictions and the non-inclusion of any information regarding uncertainty. The input data for SGP4 is a Two Line Element (TLE) set that includes information about the object and its orbit, such as satellite number, mean orbital elements, drag, ballistic coefficient, revolution number, and a time stamp. The fact that the information on Resident Space Objects is made publicly available through TLEs at space-track.org is one of the reasons for the ample use of SGP4~\cite{Vallado2013}.

\begin{figure}[!b]
    \centering
    \includegraphics[width=0.95\columnwidth]{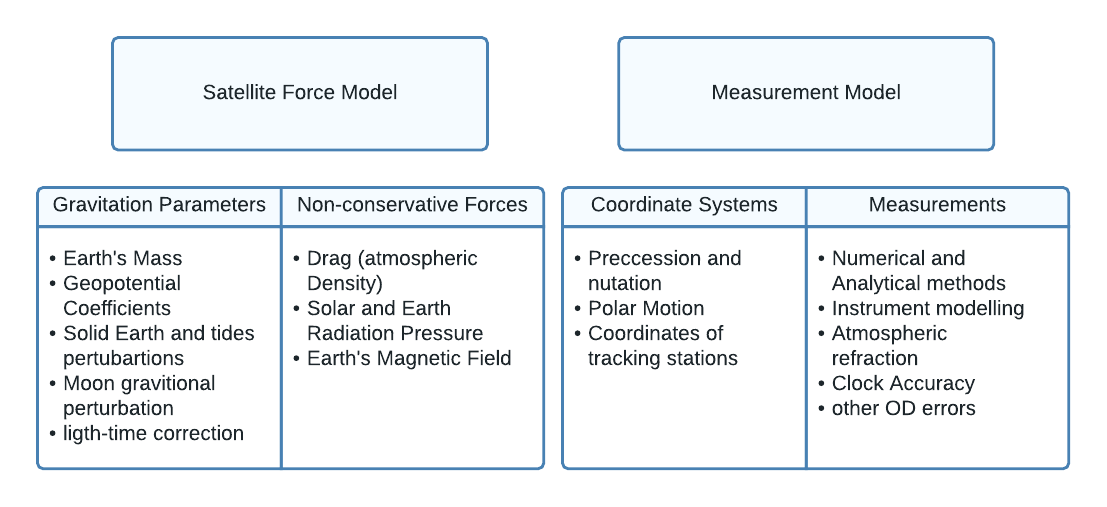}
    \caption{Error Sources in Orbital Prediction, adapted from~\citet{Luo2017}.}
    \label{fig:errors}
\end{figure}

Using the current methodology, two types of errors exist when making Orbit Propagation: satellite force model errors and measurement errors. The first error type relates to simplifications in the model, such as Earth's Gravitational harmonics model, or unknown information specific to the RSO, such as shape, attitude, or cross-sectional area. Measurement errors are all errors associated with the difference between the actual state of the satellite (such as position and velocity) and the measured state, including numerical truncation errors, as well as other navigation errors caused by clock accuracy or coordinate systems precision. In Figure~\ref{fig:errors}, we can see a summary of the error sources in Orbit Prediction, as presented by~\citet{Luo2017} and~\citet{Lei2017}.

In recent years, data-driven techniques have been used to tackle both sides of the problem, either employing machine learning to improve the model's representation of reality or mitigating the errors caused by numerical solutions and linear assumptions.
Levit and Marshall~\cite{Levit2011} showed that when using a sufficient number of TLEs, they could be used as pseudo-observations to fit a high-precision special perturbations numerical propagator. 
This method was applied to a set of satellites, and the OP error was reduced from 1.5 km/day to 0.1 km/day. This work shows that past information can be leveraged to improve Orbit Prediction, which is a key factor for any Machine Learning setup. This work was closely followed by~\citet{Bennett2012}, in which the same method was applied alongside a bias correction function. Similarly, other authors~\cite{Sang2017,SANJUAN2017254} have shown improvement in the SGP4 prediction using error correction functions based on prediction error periodicity. 

\begin{figure}[!b]
    \centering
    \includegraphics[width=0.9\textwidth]{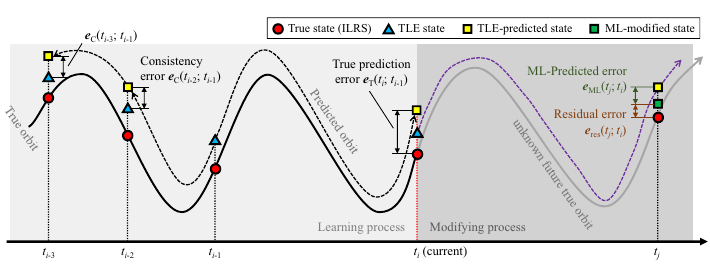}
    \caption{Illustration of the dataset structure for the ML approach to SGP4 error correction, as presented by \citet{Peng2020}.}
    \label{fig:peng2020}
\end{figure}

A second approach is to remove the physics-based propagator altogether. In~\citet{Muldoon2009}, a data-driven model was proposed to approximate the SGP4 algorithm using a polynomial fit. A different method was introduced in~\citet{Hartikainen2012} using Latent Force Models (LFMs) to combine the physical models' principles with non-parametric data-driven components. This method allows determining future orbit positions and corresponding uncertainty by assuming Gaussian process (GP) priors on unknown forces. This work was by extended Rautalin et al.~\cite{Rautalin2018}, who obtained favorable results for a set of satellite constellations in Medium Earth Orbit (MEO) and Geo-Synchronous Orbit (GEO).

Peng and Bai~\cite{Peng2018a} have extensive publications in this field. In their first paper, a Support Vector Machine (SVM) is used to learn the historical error and improve the SGP4-based prediction within the error correction framework. In a simulated catalogue, they have shown that the error-correcting model can be deployed to improve RSO's Orbit Prediction. This duo of researchers also examined the capabilities of three ML models, specifically, Artificial Neural Networks (ANN), Gaussian Processes (GP), and Support Vector Machines (SVM) in~\cite{Peng2019,Peng2018a,Peng2019a}, coming to the conclusion that the ANN had the best results, despite being more prone to overfitting. In~\citet{Peng2020}, the authors evaluated the SVM model on a real dataset and more recently proposed a data fusion approach to combine uncertainty information from the EKF OD process with the one obtained from the GP model~\cite{Peng2021}.

\begin{figure}[!b]
    \centering
    \includegraphics[width=0.6\columnwidth]{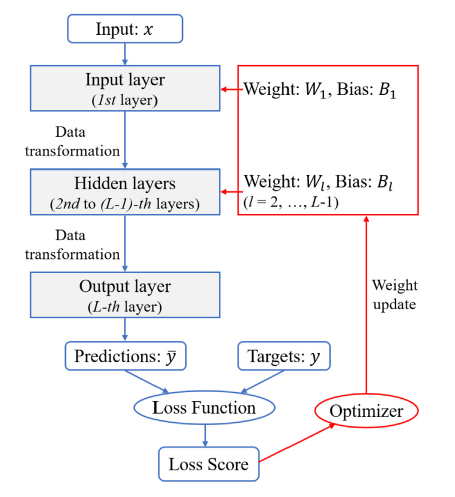}
    \caption{Implementation flow of CNN in extracting patterns from the input. Neural network topology from~\citet{Li2021}.}
    \label{fig:li2021}
\end{figure}

This hybrid approach that uses an ML algorithm to correct the error in an orbit propagator, such as the SGP4, is the most common. Various neural network architectures have been leveraged for this purpose. Under the same hybrid SGP4 approach, four independent works show that Convolutional Neural Networks (CNN)~\cite{Pihlajasalo2018}, Feed-Forward Neural Networks (FNN)~\cite{San-Juan2018}, Recurrent Neural Networks (RNN)~\cite{Curzi2022}, and Long-short Term Memory (LSTM)~\cite{Salleh2021} can be used to improve SGP4 Orbit Prediction. Using a Gradient Boosting Decision Tree (GBDT) and a CNN, Li et al.~\cite{Li2021,Li2020} improved by more than 75\% the along-track direction prediction for five satellites in LEO and GEO. All of this work has been done under the same data regimen, in particular using publicly available satellite data from the International Laser Ranging System (ILRS) as the ground truth and using a set of TLEs for a limited number of satellites as the training data. Two goals remain unaddressed in this research: the inclusion of exogenous variables in addition to TLE information and the generalization of machine learning models to unknown RSOs (referred to as type III generalization by Peng and Bai~\cite{Peng2018b}). This would allow a single Machine Learning model to be used instead of creating a separate model for each RSO.

As mentioned before, two error sources exist in Orbit Prediction, and therefore two lines of work could be developed to enhance a physics-based model. The previously shown work focus is on correcting numerical errors and observational errors that occur when using simplifying assumptions, particularly the limited SGP4 propagator. A different approach is on using ML to improve our knowledge of the Earth, space, and RSOs, thus obtaining a more exact model of reality.

\section{Thermospheric Density Mass Models}
\label{sec:ther}

Accurate prediction of the future state of Low Earth Orbit (LEO) objects requires knowledge of thermospheric mass density (the $\rho$ in~\eqref{equation:drag}). Drag is known to be the main non-conservative force affecting satellites and debris in LEO~\cite{Emmert2015}.

The state-of-the-art models currently used are the Jacchia-Bowman 2008 (JB2008)~\cite{Bowman2008}, the Drag Temperature Model 2020 (DTM-20)~\cite{Bruinsma2021}, the Naval Research Laboratory for Mass Spectrometer and Incoherent Scatter radar 2.0 model (NRLMSIS 2.0)~\cite{Emmert2021}, and corrective models such as the High Accuracy Satellite Drag Model (HASDM)~\cite{Storz2005}. These models have a long history of improvement and refinement, dating back to the 1960s.

The MSIS series of models (MSIS-77~\cite{HEDIN1}, MSIS-83~\cite{hedin2}, MSIS-86~\cite{hedin3}, MSIS-90~\cite{hedin4} and NRLMSISE-00~\cite{picone}) originally used mass spectrometer data and temperatures inferred from ground stations. Later, orbit-derived density and solar UV occultation measurements were incorporated.

The DTM series started in 1978 with the aptly named DTM-78~\cite{barlier78}, using orbit-derived density data. Subsequent models (DTM-94~\cite{Berger1998}, DTM-03~\cite{BRUINSMA2003}, DTM-09~\cite{bruinsma2012}) added accelerometer and mass spectrometer data, as well as temperature derived from incoherent scatter radar and optical airglow.

The Jachia series of models began with Jacchia-60~\cite{JACCHIA1961}. It was followed by many iterations and derived models (J65~\cite{Jacchia65}, J70~\cite{j70}, J71~\cite{71J}, MET-07~\cite{Suggs2017MarshallET} and GRAM-99~\cite{JUSTUS20041731}). More recently, the Jacchia-Bowman 2006~\cite{BOWMAN2008774} was released, incorporating new solar indices.

For a more detailed look at the history of thermospheric density mass models and a comparison of recent models, we recommend the review by~\citet{Emmert2015} and Chapter 8.6 in~\citet{Vallado2001}.

Empirical models use space weather indexes as input for forecasting and estimating local density, but the results are far from exact when used in Orbit Prediction~\cite{Vallado2014a}. On average, these models have a one-sigma accuracy of 10-15\%, depending on the model, solar activity, and location~\cite{HE201831}. The data used in their configuration limits their predictive capacity, and forecast errors of the driving features can further compromise them. For example, a 10\% error in Extreme Ultraviolet (EUV) light prediction can result in more than 200 km uncertainty for a given satellite after 7 days~\cite{Emmert2014}.

Since the 1990’s, Machine Learning has been used to forecast space weather indexes, such as the $F_{10.7}$, which is a proxy for EUV light. For example, an Artificial Neural Network (ANN) was used to predict solar flux~\cite{Williams1991}, and Time delay Neural Networks (TDNN) were used to predict geomagnetic storms~\cite{Gleisner1996PredictingGS}. There is an abundance of scientific publications on this topic~\cite{Huang2009,Yaya2017,Tobiska2000,warren2017,stevenson2022}, and a comprehensive review is available in Camporeale's comprehensive survey~\cite{Camporeale2019}.

To improve atmospheric mass density models, one approach is to use calibration data to fine-tune the parameters of the original model. This was the approach taken by~\citet{Storz2005} to develop the HASDM. Other authors have also achieved superior results with calibrated models: Doornbos et al.~\cite{DOORNBOS20081115} used Two Line Element (TLE) data to calibrate a complex physical model, halving Orbit Prediction error. \citet{2015Shi} used the same approach to calibrate the NRLMSISE-00 model during periods of increased solar activity. \citet{sang2011modification} suggested a process to calibrate any empirical model using the least squares method, which~\citet{Chen2019} found to be the most effective calibration method for reducing Orbit Prediction error. Combining multiple models can also reduce forecasting errors, as demonstrated when a Multi-model Ensemble~\cite{Elvidge2016}, which combined three physical and empirical models was able to reduce forecasting error in the training and test sets. \citet{Perez2015} used a Neural Network to combine empirical models in order to reduce density error. \citet{Chen2014} leveraged a Neural Network to calibrate models during magnetic storms, while \citet{Zhang2021} and \citet{GAO2020273} used LSTM and Gaussian Processes, respectively, to calibrate the original models.

\begin{figure}[!b]
    \centering
    \includegraphics[width=0.9\columnwidth]{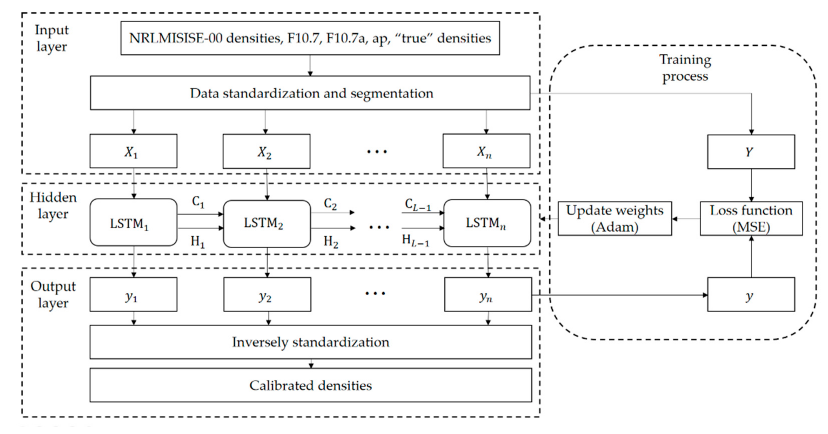}
    \caption{The structure of the LSTM-NRL calibration model in \citet{Zhang2021}.}
    \label{fig:zHANG}
\end{figure}

Recently, a different approach has emerged, using Reduced Order Models (ROM) as a quasi-physical model. Mehta et al.~\cite{Mehta2018} used a ROM to approximate the TIE-GCM complex physical model and simulated orbit ephemerides for model calibration. ROMs reduce the high-dimensional space to a lower dimension while preserving most information. This is usually done using Proper Orthogonal Decomposition (POD), also known as Principal Component Analysis (PCA), in the Machine Learning world.

This work was extended by using accelerometer-derived density estimates from the CHAMP and GOCE satellites~\cite{Mehta_2018}, as well as TLE data~\cite{Gondelach2020}. Gondelach and Linares~\cite{Gondelach2021} showed that a ROM model with TLE-estimated densities provides more precise Orbit Determination than the NRLMSISE-00 and JB2008 models, particularly in the along-track coordinate. Turner et al.~\cite{Turner2020} improved this ROM approach by using an autoencoder for dimensionality reduction. Nateghi et al.~\cite{Nateghi2021} further refined it by using Sparse Identification of Nonlinear Dynamical systems (SINDy) to derive an explicit non-linear differential equation that defined the atmospheric density in the encoded (low-dimension) space.

In 2017, Mehta et al.~\cite{Mehta2017} published a global density estimate derived from the GRACE and CHAMP satellites using a physical model with high-fidelity drag coefficient modeling. This data was then used to develop an LSTM-based Neural Network for global density prediction, with promising results~\cite{George2021}. Bonasera et al.~\cite{Bonasera2021} used two Machine Learning techniques to derive uncertainty over the estimated density: Monte-Carlo Dropout and Deep Ensemble. Both techniques use sampling to obtain a probability distribution of the estimated quantity. Monte-Carlo Dropout ``turns off'' a percentage of the neurons in a Neural Network for each sample~\cite{GoodBengCour16}, while Deep Ensemble combines the output of a series of Neural Networks to obtain an estimated mean and uncertainty of the prediction~\cite{Steven2018}.

\begin{figure}[!b]
    \centering
    \includegraphics[width=0.95\columnwidth]{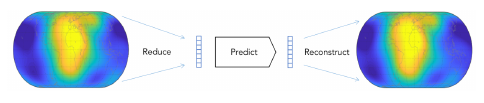}
    \caption{Reduced Order Model (ROM) of thermospheric density conceptual representation. Where the prediction step is performed on a low dimension space and then the thermospheric density is reconstructed, as shown in \citet{Nateghi2021}.}
    \label{fig:nat}
\end{figure}

The same group of researchers also developed a method to forecast the input features for empirical models with uncertainty estimation, combining solar data images with time-series information~\cite{Benson2021}. In 2020, SET (Space Environment Technologies) made public 20 years of density data derived from HASDM, the model used by the United States Air Force~\cite{Tobiska2021}. This data was used to train a NN to replicate the HASDM model~\cite{Licata2022}, using the same input, and to add uncertainty information through Monte-Carlo Dropout. Licata et al.~\cite{Licata2022a} further advanced this work by using Bayesian Neural Networks on the global HASDM dataset and the local CHAMP dataset. Bayesian Neural Networks have recently gained more attention in the machine learning world, as they allow for a more principled understanding of uncertainty when compared to Monte-Carlo Dropout or Ensembles. They replace the neurons in a network, which are unknown parameters, or weights, with random variables, with the training process learning the probability distribution function of each neuron~\cite{Goan_2020}.

\section{Future Challenges and Conclusion}

In this section, we discuss some of the current and future lines of research in the field. We observe a shift from linear and physical models to more data-driven, non-linear models. Linear models were first used in the 1960s~\cite{Smith1962}, but computational limitations have since been surpassed. The Gaussian assumption used in the Extended Kalman Filter (EKF) is now being questioned~\cite{Poore2016}, and new approaches using Gaussian Mixture Models (GMM)~\cite{Vittaldev2015}, and the Gauss Von Mises (GVM) distribution~\cite{Horwood2014} are promising. However, two main drawbacks remain: determining the exact number of Gaussian distributions necessary for accurate orbit determination and calculating the Probability of Collision (PoC). The GMM was evaluated for PoC using Monte Carlo simulation in~\citet{Vittaldev2015}, but this is too time-consuming for general use. \citet{Hall2021} generalized the PoC equation to any distribution and hypothesized the use of GMMs.

Most machine learning in this area is focused on density estimation, with limited use of Deep Learning techniques, with the exception of~\citet{Chipade2021}, who used an EKF-ANN hybrid. Many Deep Learning architectures have been used for Orbit Prediction, most of them with a focus on improving upon the SGP4 propagator. However, uncertainty in the ML-OP scheme is necessary for industrial use. Bayesian counterparts of Neural Networks, which predict a distribution, are being explored, such as~\citet{Peng2021}, who combined Gaussian Processes (GP) uncertainty with EKF-derived uncertainty, and~\citet{Curzi2022}, who used a Deep Ensemble to perform variance estimation.

Research is also being done on estimating unknown physical properties of RSOs. For example,~\citet{GUTHRIE2022} used Siamese CNNs to estimate attitude, and~\citet{Sharma} and~\citet{chen2019satellite} used CNNs combined with a 3D image generator to determine the pose of satellites. \citet{Furfaro2019} leveraged deep learning techniques to determine RSOs shape using light curves data. Meta-Learning Models~\cite{Furfaro2019a} are being explored to address the lack of labeled data.

Atmospheric density and air drag forces are considered in any current force model, but the current state-of-the-art models (e.g., JB2008, DTM) have 10 to 15\% 1-sigma accuracy~\cite{Vallado2014a}. Recent papers~\cite{Licata2022,Bonasera2021,Nateghi2021} have proposed replacing these empirical models with modern deep learning models. Grey-box models, such as the one proposed by~\citet{Nateghi2021}, are becoming more common since they are easier to evaluate and can implement physics-aware restrictions.

%----------

\begin{figure}[!tb]
    \centering
    \includegraphics[width=0.9\columnwidth,scale=0.5]{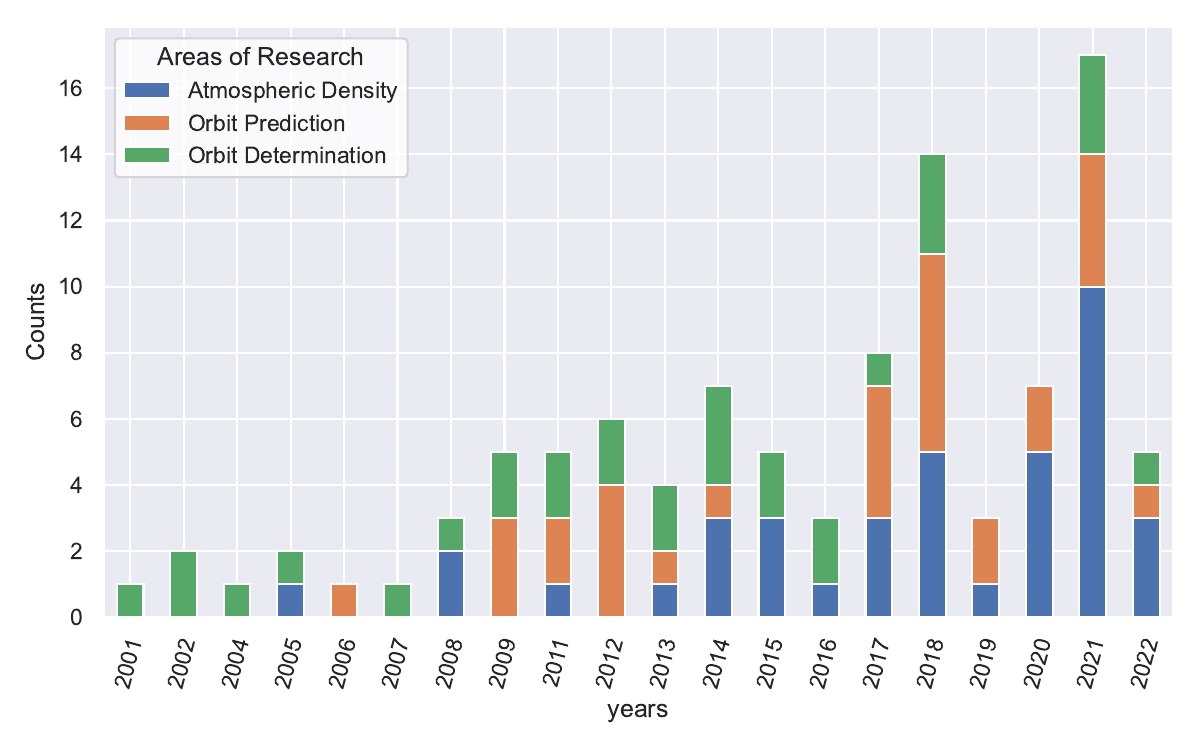}
    \caption{Number of publications between 2000 and 2022 cited in this review, with applications divided by three categories: Orbital Determination, Orbital Prediction and Atmospheric Density Mass models.}
    \label{fig:y}
\end{figure}

%In conclusion, we see a field where classical algorithms can eventually be enhanced by ML, but that will only occur if we use machine learning techniques to represent better and understand the underlying physical forces. Across the board, the predictions will become more focused on probability distributions than point-wise estimations. The push to create grey-box models, simpler physics-aware algorithms that can be understood by physicists and machine learning specialists alike, will gain prominence, as that is the real problem --- How can we use machine learning to gain a better understanding of a system or, in this case, the physical forces that govern space?

In conclusion, ML techniques can enhance and improve classical algorithms to help us better represent and understand the underlying physical phenomena. With the help of ML, predictions may become more focused on probability distributions rather than point-wise estimations, and the development of grey-box models - models that are simpler and more comprehensible to both space and ML specialists - can become increasingly important. This is the real challenge: how can ML be used to understand a system better, and in this case, the physical forces that govern space? In order to do this, we need to be able to combine the knowledge and expertise of both ML and space experts to create an environment that promotes collaboration and understanding. This can be achieved through the development of models which are mutually comprehensible and the introduction of ML-based tools that can help to bridge the gap between the two disciplines. Ultimately, this will help to create a more unified and practical approach to modeling physical processes.

\section*{Acknowledgements}
This work was partially supported by FCT/PT Space under the PhD grant PRT/BD/153601/2021 and the strategic project NOVA LINCS (UIDB/ 04516/2020).  Funded by the European Union Project (TARDIS, 101093006). Views and opinions expressed are however those of the author(s) only and do not necessarily reflect those of the European Union. Neither the European Union nor the granting authority can be held responsible for them.

\noindent\includegraphics[width=2cm]{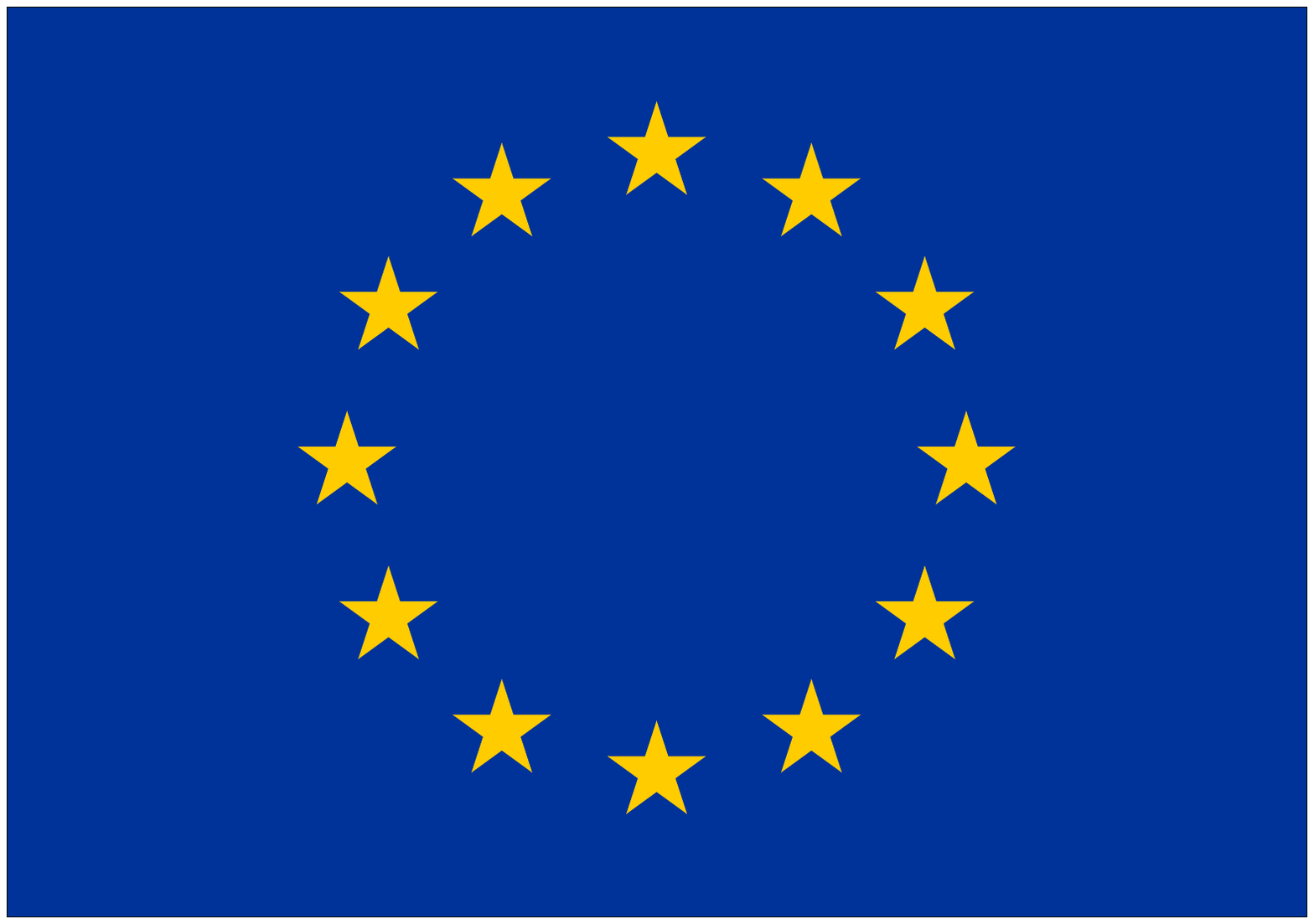}

%% The Appendices part is started with the command \appendix;
%% appendix sections are then done as normal sections
%%\appendix

%%\section{Sample Appendix Section}
%%\label{sec:sample:appendix}
%Lorem ipsum dolor sit amet, consectetur adipiscing elit, sed do eiusmod tempor section \ref{sec:sample1} incididunt ut labore et dolore magna aliqua. Ut enim ad minim veniam, quis nostrud exercitation ullamco laboris nisi ut aliquip ex ea commodo consequat. Duis aute irure dolor in reprehenderit in voluptate velit esse cillum dolore eu fugiat nulla pariatur. Excepteur sint occaecat cupidatat non proident, sunt in culpa qui officia deserunt mollit anim id est laborum.

%% If you have bibdatabase file and want bibtex to generate the
%% bibitems, please use
%%
 \bibliographystyle{elsarticle-num-names} 
 \bibliography{cas-refs}

\begin{thebibliography}{146}
\expandafter\ifx\csname natexlab\endcsname\relax\def\natexlab#1{#1}\fi
\providecommand{\url}[1]{\texttt{#1}}
\providecommand{\href}[2]{#2}
\providecommand{\path}[1]{#1}
\providecommand{\DOIprefix}{doi:}
\providecommand{\ArXivprefix}{arXiv:}
\providecommand{\URLprefix}{URL: }
\providecommand{\Pubmedprefix}{pmid:}
\providecommand{\doi}[1]{\href{http://dx.doi.org/#1}{\path{#1}}}
\providecommand{\Pubmed}[1]{\href{pmid:#1}{\path{#1}}}
\providecommand{\bibinfo}[2]{#2}
\ifx\xfnm\relax \def\xfnm[#1]{\unskip,\space#1}\fi
%Type = Misc
\bibitem[{{ESA's Space Debris Office}(2022)}]{esa_2021}
\bibinfo{author}{{ESA's Space Debris Office}}, \bibinfo{title}{Space debris by
  the numbers}, \bibinfo{year}{2022}. \URLprefix
  \url{https://www.esa.int/Safety_Security/Space_Debris/Space_debris_by_the_numbers},
  \bibinfo{note}{accessed May 10, 2022}.
%Type = Book
\bibitem[{Vallado(2001)}]{Vallado2001}
\bibinfo{author}{D.~A. Vallado}, \bibinfo{title}{{Fundamentals of Astrodynamics
  and Applications}}, volume~\bibinfo{volume}{12}, \bibinfo{edition}{4th} ed.,
  \bibinfo{publisher}{Springer Science \& Business Media},
  \bibinfo{year}{2001}.
%Type = Incollection
\bibitem[{Curtis(2014)}]{curtis2014}
\bibinfo{author}{H.~D. Curtis},
\newblock \bibinfo{title}{Preliminary orbit determination},
\newblock in: \bibinfo{booktitle}{Orbital Mechanics for Engineering Students
  (Third Edition)}, \bibinfo{edition}{third edition} ed.,
  \bibinfo{publisher}{Butterworth-Heinemann}, \bibinfo{address}{Boston},
  \bibinfo{year}{2014}, pp. \bibinfo{pages}{239--298}.
%Type = Techreport
\bibitem[{Smith et~al.(1962)Smith, Schmidt, Mcgee, and Field}]{Smith1962}
\bibinfo{author}{G.~L. Smith}, \bibinfo{author}{S.~F. Schmidt},
  \bibinfo{author}{L.~A. Mcgee}, \bibinfo{author}{M.~Field},
  \bibinfo{title}{{Application of statistical filter theory to the optimal
  estimation of position and velocity on board a circumlunar vehicle}},
  \bibinfo{type}{Technical Report}, NASA, \bibinfo{year}{1962}.
  \bibinfo{note}{R-135}.
%Type = Article
\bibitem[{Julier and Uhlmann(1997)}]{Julier1997}
\bibinfo{author}{S.~J. Julier}, \bibinfo{author}{J.~K. Uhlmann},
\newblock \bibinfo{title}{{New extension of the Kalman filter to nonlinear
  systems}},
\newblock \bibinfo{journal}{Signal Processing, Sensor Fusion, and Target
  Recognition VI} \bibinfo{volume}{3068} (\bibinfo{year}{1997})
  \bibinfo{pages}{182 -- 193}.
%Type = Incollection
\bibitem[{Einicke(2012)}]{Einick}
\bibinfo{author}{G.~A. Einicke},
\newblock \bibinfo{title}{Nonlinear prediction, filtering and smoothing},
\newblock in: \bibinfo{booktitle}{Smoothing, Filtering and Prediction -
  Estimating The Past, Present and Future}, \bibinfo{edition}{2nd} ed.,
  \bibinfo{publisher}{IntechOpen}, \bibinfo{address}{Rijeka},
  \bibinfo{year}{2012}, pp. \bibinfo{pages}{245--275}.
%Type = Incollection
\bibitem[{Tapley et~al.(2004)Tapley, Schutz, and Born}]{tapley2004}
\bibinfo{author}{B.~D. Tapley}, \bibinfo{author}{B.~E. Schutz},
  \bibinfo{author}{G.~H. Born},
\newblock \bibinfo{title}{Fundamentals of orbit determination},
\newblock in: \bibinfo{editor}{B.~D. Tapley}, \bibinfo{editor}{B.~E. Schutz},
  \bibinfo{editor}{G.~H. Born} (Eds.), \bibinfo{booktitle}{Statistical Orbit
  Determination}, \bibinfo{publisher}{Academic Press},
  \bibinfo{address}{Burlington}, \bibinfo{year}{2004}, pp.
  \bibinfo{pages}{159--284}.
%Type = Book
\bibitem[{Montenbruck and Gill(2005)}]{Montenbruck2005}
\bibinfo{author}{O.~Montenbruck}, \bibinfo{author}{E.~Gill},
  \bibinfo{title}{Satellite Orbits: Models, Methods, and Applications},
  volume~\bibinfo{volume}{1}, \bibinfo{publisher}{Springer Berlin, Heidelberg},
  \bibinfo{year}{2005}.
  \DOIprefix\doi{https://doi.org/10.1007/978-3-642-58351-3}.
%Type = Book
\bibitem[{Bate et~al.(1971)Bate, Mueller, and White}]{bate1971fundamentals}
\bibinfo{author}{R.~Bate}, \bibinfo{author}{D.~Mueller},
  \bibinfo{author}{J.~White}, \bibinfo{title}{Fundamentals of Astrodynamics},
  Dover Books on Aeronautical Engineering Series, \bibinfo{publisher}{Dover
  Publications}, \bibinfo{year}{1971}.
%Type = Inproceedings
\bibitem[{Uphoff(1972)}]{UPHOFF1972}
\bibinfo{author}{C.~Uphoff},
\newblock \bibinfo{title}{{Numerical Averaging in Orbit Prediction}},
\newblock in: \bibinfo{booktitle}{Astrodynamics Conference},
  volume~\bibinfo{volume}{11}, \bibinfo{publisher}{AIAA},
  \bibinfo{address}{Palo Alto,CA}, \bibinfo{year}{1972}, pp.
  \bibinfo{pages}{1512--1516}.
%Type = Techreport
\bibitem[{Poore et~al.(2016)Poore, Aristoff, and Horwood}]{Poore2016}
\bibinfo{author}{A.~B. Poore}, \bibinfo{author}{J.~M. Aristoff},
  \bibinfo{author}{J.~T. Horwood}, \bibinfo{title}{{Covariance and Uncertainty
  Realism in Space Surveillance and Tracking Working Group on Covariance
  Realism}}, \bibinfo{type}{Technical Report}, Air Force Space Command
  Astrodynamics Innovation Committee, \bibinfo{year}{2016}.
%Type = Article
\bibitem[{Storz et~al.(2005)Storz, Bowman, Branson, Casali, and
  Tobiska}]{Storz2005}
\bibinfo{author}{M.~F. Storz}, \bibinfo{author}{B.~R. Bowman},
  \bibinfo{author}{J.~I. Branson}, \bibinfo{author}{S.~J. Casali},
  \bibinfo{author}{W.~K. Tobiska},
\newblock \bibinfo{title}{{High accuracy satellite drag model (HASDM)}},
\newblock \bibinfo{journal}{Advances in Space Research} \bibinfo{volume}{36}
  (\bibinfo{year}{2005}) \bibinfo{pages}{2497--2505}.
%Type = Article
\bibitem[{Horwood et~al.(2011)Horwood, Aragon, and Poore}]{Horwood2011}
\bibinfo{author}{J.~T. Horwood}, \bibinfo{author}{N.~D. Aragon},
  \bibinfo{author}{A.~B. Poore},
\newblock \bibinfo{title}{Gaussian sum filters for space surveillance: Theory
  and simulations},
\newblock \bibinfo{journal}{Journal of Guidance, Control, and Dynamics}
  \bibinfo{volume}{34} (\bibinfo{year}{2011}) \bibinfo{pages}{1839--1851}.
%Type = Article
\bibitem[{Pavlis et~al.(2012)Pavlis, Holmes, Kenyon, and Factor}]{pavlis2012}
\bibinfo{author}{N.~K. Pavlis}, \bibinfo{author}{S.~A. Holmes},
  \bibinfo{author}{S.~C. Kenyon}, \bibinfo{author}{J.~K. Factor},
\newblock \bibinfo{title}{{The development and evaluation of the Earth
  Gravitational Model 2008 (EGM2008)}},
\newblock \bibinfo{journal}{Journal of Geophysical Research: Solid Earth}
  \bibinfo{volume}{117} (\bibinfo{year}{2012}).
%Type = Inproceedings
\bibitem[{Vallado et~al.(2006)Vallado, Crawford, Hujsak, and
  Kelso}]{Vallado2006}
\bibinfo{author}{D.~A. Vallado}, \bibinfo{author}{P.~Crawford},
  \bibinfo{author}{R.~Hujsak}, \bibinfo{author}{T.~S. Kelso},
\newblock \bibinfo{title}{{Revisiting spacetrack report \#3}},
\newblock in: \bibinfo{booktitle}{{AIAA}/{AAS} Astrodynamics Specialist
  Conference and Exhibit}, volume~\bibinfo{volume}{3},
  \bibinfo{publisher}{AIAA}, \bibinfo{year}{2006}, pp.
  \bibinfo{pages}{1984--2071}.
%Type = Inbook
\bibitem[{Doornbos(2012)}]{Doornbos2012}
\bibinfo{author}{E.~Doornbos}, \bibinfo{title}{Satellite Dynamics and
  Non-Gravitational Force Modelling}, \bibinfo{publisher}{Springer Berlin
  Heidelberg}, \bibinfo{address}{Berlin, Heidelberg}, \bibinfo{year}{2012}, pp.
  \bibinfo{pages}{59--89}.
%Type = Article
\bibitem[{Zhong~Luo and Yang(2017)}]{Luo2017}
\bibinfo{author}{Y.~Zhong~Luo}, \bibinfo{author}{Z.~Yang},
\newblock \bibinfo{title}{{A review of uncertainty propagation in orbital
  mechanics}},
\newblock \bibinfo{journal}{Progress in Aerospace Sciences}
  \bibinfo{volume}{89} (\bibinfo{year}{2017}) \bibinfo{pages}{23--39}.
%Type = Book
\bibitem[{Milani and Gronchi(2009)}]{Milani2009}
\bibinfo{author}{A.~Milani}, \bibinfo{author}{G.~F. Gronchi},
  \bibinfo{title}{{Theory of orbit determination}},
  \bibinfo{publisher}{Cambridge University Press}, \bibinfo{year}{2009}.
%Type = Inproceedings
\bibitem[{Mortlock and Kassas(2021)}]{Mortlock2021}
\bibinfo{author}{T.~Mortlock}, \bibinfo{author}{Z.~M. Kassas},
\newblock \bibinfo{title}{Assessing machine learning for {LEO} satellite orbit
  determination in simultaneous tracking and navigation},
\newblock in: \bibinfo{booktitle}{2021 {Institute of Electric and Electronics
  Engineers (IEEE)} Aerospace Conference}, \bibinfo{year}{2021}, pp.
  \bibinfo{pages}{1--8}.
%Type = Article
\bibitem[{Krener(2007)}]{Krener2007}
\bibinfo{author}{A.~J. Krener},
\newblock \bibinfo{title}{{The Convergence of the Extended Kalman Filter}},
\newblock \bibinfo{journal}{Directions in Mathematical Systems Theory and
  Optimization}  (\bibinfo{year}{2007}) \bibinfo{pages}{173--182}.
%Type = Article
\bibitem[{Kalman(1960)}]{kalman1960}
\bibinfo{author}{R.~E. Kalman},
\newblock \bibinfo{title}{A new approach to linear filtering and prediction
  problems},
\newblock \bibinfo{journal}{Journal of Basic Engineering} \bibinfo{volume}{82}
  (\bibinfo{year}{1960}) \bibinfo{pages}{35--45}.
%Type = Article
\bibitem[{Ribeiro and Ribeiro(2004)}]{Ribeiro2004}
\bibinfo{author}{M.~Ribeiro}, \bibinfo{author}{I.~Ribeiro},
\newblock \bibinfo{title}{Kalman and extended kalman filters: Concept,
  derivation and properties},
\newblock \bibinfo{journal}{Institute for Systems and Robotics}
  \bibinfo{volume}{43} (\bibinfo{year}{2004}) \bibinfo{pages}{46}.
%Type = Book
\bibitem[{Gelb et~al.(1974)Gelb, Kasper, Nash, Price, and
  Sutherland}]{Gelb1974}
\bibinfo{editor}{A.~Gelb}, \bibinfo{editor}{J.~F. Kasper},
  \bibinfo{editor}{R.~A. Nash}, \bibinfo{editor}{C.~F. Price},
  \bibinfo{editor}{A.~A. Sutherland} (Eds.), \bibinfo{title}{Applied Optimal
  Estimation}, \bibinfo{publisher}{MIT Press}, \bibinfo{address}{Cambridge,
  MA}, \bibinfo{year}{1974}.
%Type = Book
\bibitem[{Kim(2011)}]{Kim2011}
\bibinfo{author}{P.~Kim}, \bibinfo{title}{{Kalman Filter for Beginners: With
  {MATLAB} examples}}, \bibinfo{publisher}{Createspace Independent Publishing
  Platform}, \bibinfo{address}{North Charleston, SC}, \bibinfo{year}{2011}.
%Type = Incollection
\bibitem[{Kim and Bang(2018)}]{Bang18}
\bibinfo{author}{Y.~Kim}, \bibinfo{author}{H.~Bang},
\newblock \bibinfo{title}{{Introduction to Kalman Filter and Its
  Applications}},
\newblock in: \bibinfo{editor}{F.~Govaers} (Ed.),
  \bibinfo{booktitle}{Introduction and Implementations of the Kalman Filter},
  \bibinfo{publisher}{IntechOpen}, \bibinfo{address}{Rijeka},
  \bibinfo{year}{2018}, p.~\bibinfo{pages}{19}.
%Type = Phdthesis
\bibitem[{Lee(2005)}]{Lee2005}
\bibinfo{author}{D.-J. Lee}, \bibinfo{title}{{Nonlinear bayesian filtering with
  applications to estimation and navigation}}, Ph.D. thesis, Texas A\&M
  University, \bibinfo{address}{College Station, TX}, \bibinfo{year}{2005}.
  \bibinfo{note}{ProQuest Ebrary}.
%Type = Inproceedings
\bibitem[{Payne and Marrs(2004)}]{payne2004}
\bibinfo{author}{O.~Payne}, \bibinfo{author}{A.~Marrs},
\newblock \bibinfo{title}{An unscented particle filter for gmti tracking},
\newblock in: \bibinfo{booktitle}{2004 Institute of Electric and Electronics
  Engineers (IEEE) Aerospace Conference Proceedings},
  volume~\bibinfo{volume}{3}, \bibinfo{year}{2004}, pp.
  \bibinfo{pages}{1869--1875}.
%Type = Article
\bibitem[{Arulampalam et~al.(2002)Arulampalam, Maskell, Gordon, and
  Clapp}]{Arulampalam2002}
\bibinfo{author}{M.~S. Arulampalam}, \bibinfo{author}{S.~Maskell},
  \bibinfo{author}{N.~Gordon}, \bibinfo{author}{T.~Clapp},
\newblock \bibinfo{title}{{A Tutorial on Particle Filters for Online
  Nonlinear/Nongaussian Bayesian Tracking}},
\newblock \bibinfo{journal}{Institute of Electric and Electronics Engineers
  (IEEE) Transactions on Signal Processing} \bibinfo{volume}{50}
  (\bibinfo{year}{2002}) \bibinfo{pages}{723--737}.
%Type = Article
\bibitem[{Ning et~al.(2012)Ning, Ma, Peng, Quan, and Fang}]{Ning2012}
\bibinfo{author}{X.~Ning}, \bibinfo{author}{X.~Ma}, \bibinfo{author}{C.~Peng},
  \bibinfo{author}{W.~Quan}, \bibinfo{author}{J.~Fang},
\newblock \bibinfo{title}{{Analysis of filtering methods for satellite
  autonomous orbit determination using celestial and geomagnetic measurement}},
\newblock \bibinfo{journal}{Mathematical Problems in Engineering}
  \bibinfo{volume}{2012} (\bibinfo{year}{2012}).
%Type = Inproceedings
\bibitem[{Hartikainen et~al.(2012)Hartikainen, Sepp{\"{a}}nen, and
  S{\"{a}}rkk{\"{a}}}]{Hartikainen2012}
\bibinfo{author}{J.~Hartikainen}, \bibinfo{author}{M.~Sepp{\"{a}}nen},
  \bibinfo{author}{S.~S{\"{a}}rkk{\"{a}}},
\newblock \bibinfo{title}{{State-Space Inference for Non-Linear Latent Force
  Models with Application to Satellite Orbit Prediction}},
\newblock in: \bibinfo{booktitle}{Proceedings of the 29th International
  Conference on Machine Learning}, \bibinfo{publisher}{ICML},
  \bibinfo{address}{Edinburgh, Scotland}, \bibinfo{year}{2012}, pp.
  \bibinfo{pages}{903--910}.
%Type = Inproceedings
\bibitem[{Álvarez et~al.(2009)Álvarez, Luengo, and Lawrence}]{alvarez2019}
\bibinfo{author}{M.~Álvarez}, \bibinfo{author}{D.~Luengo},
  \bibinfo{author}{N.~D. Lawrence},
\newblock \bibinfo{title}{Latent force models},
\newblock in: \bibinfo{editor}{D.~van Dyk}, \bibinfo{editor}{M.~Welling}
  (Eds.), \bibinfo{booktitle}{Proceedings of the Twelth International
  Conference on Artificial Intelligence and Statistics},
  volume~\bibinfo{volume}{5} of \textit{\bibinfo{series}{Proceedings of Machine
  Learning Research}}, \bibinfo{publisher}{PMLR}, \bibinfo{address}{Hilton
  Clearwater Beach Resort, Clearwater Beach, Florida USA},
  \bibinfo{year}{2009}, pp. \bibinfo{pages}{9--16}.
%Type = Article
\bibitem[{Särkkä et~al.(2019)Särkkä, Álvarez, and Lawrence}]{sarkka2019}
\bibinfo{author}{S.~Särkkä}, \bibinfo{author}{M.~A. Álvarez},
  \bibinfo{author}{N.~D. Lawrence},
\newblock \bibinfo{title}{Gaussian process latent force models for learning and
  stochastic control of physical systems},
\newblock \bibinfo{journal}{IEEE Transactions on Automatic Control}
  \bibinfo{volume}{64} (\bibinfo{year}{2019}) \bibinfo{pages}{2953--2960}.
%Type = Article
\bibitem[{Raissi et~al.(2019)Raissi, Perdikaris, and
  Karniadakis}]{pinnsProposed}
\bibinfo{author}{M.~Raissi}, \bibinfo{author}{P.~Perdikaris},
  \bibinfo{author}{G.~E. Karniadakis},
\newblock \bibinfo{title}{Physics-informed neural networks: A deep learning
  framework for solving forward and inverse problems involving nonlinear
  partial differential equations},
\newblock \bibinfo{journal}{Journal of Computational Physics}
  \bibinfo{volume}{378} (\bibinfo{year}{2019}) \bibinfo{pages}{686--707}.
%Type = Inproceedings
\bibitem[{Ghilardi et~al.(2022)Ghilardi, Scorsoglio, and Furfaro}]{ghilardi}
\bibinfo{author}{L.~Ghilardi}, \bibinfo{author}{A.~Scorsoglio},
  \bibinfo{author}{R.~Furfaro},
\newblock \bibinfo{title}{Orbit determination with maneuver estimation in
  cislunar environment va physics informed neural networks},
\newblock in: \bibinfo{booktitle}{2022 AAS/AIAA Astrodynamics Specialist
  Conference}, \bibinfo{address}{Charlotte, USA}, \bibinfo{year}{2022},
  p.~\bibinfo{pages}{13}.
%Type = Inproceedings
\bibitem[{Scorsoglio et~al.(2023)Scorsoglio, D’Ambrosio, Ghilardi, Furfaro,
  and Reddy}]{scorsoglio2023physics}
\bibinfo{author}{A.~Scorsoglio}, \bibinfo{author}{A.~D’Ambrosio},
  \bibinfo{author}{L.~Ghilardi}, \bibinfo{author}{R.~Furfaro},
  \bibinfo{author}{V.~Reddy},
\newblock \bibinfo{title}{Physics-informed orbit determination for cislunar
  space applications},
\newblock in: \bibinfo{booktitle}{Proceedings of the Advanced Maui Optical and
  Space Surveillance (AMOS) Technologies Conference}, \bibinfo{address}{Wailea,
  HI}, \bibinfo{year}{2023}, p.~\bibinfo{pages}{10}.
%Type = Techreport
\bibitem[{Sharma and Cutler(2015)}]{Sharma2015}
\bibinfo{author}{S.~Sharma}, \bibinfo{author}{J.~W. Cutler},
  \bibinfo{title}{{Robust Classification and Orbit Determination : A Learning
  Theoretic Approach}}, \bibinfo{type}{Technical Report}, NASA Interplanetary
  Network Progress Report 42-203, \bibinfo{year}{2015}.
%Type = Article
\bibitem[{Szab{\'o} et~al.(2016)Szab{\'o}, Sriperumbudur, P{\'o}czos, and
  Gretton}]{szabo2016learning}
\bibinfo{author}{Z.~Szab{\'o}}, \bibinfo{author}{B.~K. Sriperumbudur},
  \bibinfo{author}{B.~P{\'o}czos}, \bibinfo{author}{A.~Gretton},
\newblock \bibinfo{title}{Learning theory for distribution regression},
\newblock \bibinfo{journal}{The Journal of Machine Learning Research}
  \bibinfo{volume}{17} (\bibinfo{year}{2016}) \bibinfo{pages}{5272--5311}.
%Type = Phdthesis
\bibitem[{Sharma(2018)}]{Sharma2018}
\bibinfo{author}{S.~Sharma}, \bibinfo{title}{{Machine Learning Applications in
  Spacecraft State and Environment Estimation}}, Ph.D. thesis, University of
  Michigan, \bibinfo{address}{Ann Arbor, MI}, \bibinfo{year}{2018}.
  \bibinfo{note}{ProQuest Ebrary}.
%Type = Article
\bibitem[{Jiang(2021)}]{Jiang2021}
\bibinfo{author}{C.~Jiang},
\newblock \bibinfo{title}{{An orbit determination method of spacecraft based on
  distribution regression}},
\newblock \bibinfo{journal}{Open Astronomy} \bibinfo{volume}{30}
  (\bibinfo{year}{2021}) \bibinfo{pages}{159--167}.
%Type = Article
\bibitem[{Terejanu et~al.(2008)Terejanu, Singla, Singh, and
  Scott}]{Terejanu2008}
\bibinfo{author}{G.~Terejanu}, \bibinfo{author}{P.~Singla},
  \bibinfo{author}{T.~Singh}, \bibinfo{author}{P.~D. Scott},
\newblock \bibinfo{title}{{Uncertainty Propagation for Nonlinear Dynamic
  Systems using Gaussian Mixture Models}},
\newblock \bibinfo{journal}{Journal of Guidance, Control, and Dynamics}
  \bibinfo{volume}{31} (\bibinfo{year}{2008}) \bibinfo{pages}{1623--1633}.
%Type = Article
\bibitem[{DeMars et~al.(2013)DeMars, Bishop, and Jah}]{DeMars2013}
\bibinfo{author}{K.~DeMars}, \bibinfo{author}{R.~Bishop},
  \bibinfo{author}{M.~Jah},
\newblock \bibinfo{title}{Entropy-based approach for uncertainty propagation of
  nonlinear dynamical systems},
\newblock \bibinfo{journal}{Journal of Guidance, Control, and Dynamics}
  \bibinfo{volume}{36} (\bibinfo{year}{2013}) \bibinfo{pages}{1047--1057}.
%Type = Article
\bibitem[{Alspach and Sorenson(1972)}]{Alspach1972}
\bibinfo{author}{D.~L. Alspach}, \bibinfo{author}{H.~W. Sorenson},
\newblock \bibinfo{title}{{Nonlinear Bayesian Estimation using Gaussian Sum
  Approximations}},
\newblock \bibinfo{journal}{Institute of Electric and Electronics Engineers
  (IEEE) Transactions on Automatic Control} \bibinfo{volume}{17}
  (\bibinfo{year}{1972}) \bibinfo{pages}{439--448}.
%Type = Book
\bibitem[{Goodfellow et~al.(2016)Goodfellow, Bengio, and
  Courville}]{GoodBengCour16}
\bibinfo{author}{I.~J. Goodfellow}, \bibinfo{author}{Y.~Bengio},
  \bibinfo{author}{A.~Courville}, \bibinfo{title}{Deep Learning},
  \bibinfo{publisher}{MIT Press}, \bibinfo{address}{Cambridge, MA, USA},
  \bibinfo{year}{2016}.
%Type = Article
\bibitem[{Vishwajeet et~al.(2014)Vishwajeet, Singla, and Jah}]{Vishwajeet2014}
\bibinfo{author}{K.~Vishwajeet}, \bibinfo{author}{P.~Singla},
  \bibinfo{author}{M.~Jah},
\newblock \bibinfo{title}{{Nonlinear Uncertainty Propagation for Perturbed
  Two-body Orbits}},
\newblock \bibinfo{journal}{Journal of Guidance, Control, and Dynamics}
  \bibinfo{volume}{37} (\bibinfo{year}{2014}) \bibinfo{pages}{1415--1425}.
%Type = Article
\bibitem[{Terejanu et~al.(2011)Terejanu, Singla, Singh, and
  Scott}]{Terejanu2011}
\bibinfo{author}{G.~Terejanu}, \bibinfo{author}{P.~Singla},
  \bibinfo{author}{T.~Singh}, \bibinfo{author}{P.~D. Scott},
\newblock \bibinfo{title}{{Adaptive Gaussian Sum Filter for Nonlinear Bayesian
  Estimation}},
\newblock \bibinfo{journal}{Institute of Electric and Electronics Engineers
  (IEEE) Transactions on Automatic Control} \bibinfo{volume}{56}
  (\bibinfo{year}{2011}) \bibinfo{pages}{2151--2156}.
%Type = Phdthesis
\bibitem[{Vittaldev(2015)}]{Vittaldev2015}
\bibinfo{author}{V.~Vittaldev}, \bibinfo{title}{{Uncertainty Propagation and
  Conjunction Assessment for Resident Space Objects}}, Ph.D. thesis, University
  of Texas, \bibinfo{address}{Austin, TX}, \bibinfo{year}{2015}.
%Type = Article
\bibitem[{Vittaldev and Russell(2016)}]{Vittaldev2016}
\bibinfo{author}{V.~Vittaldev}, \bibinfo{author}{R.~P. Russell},
\newblock \bibinfo{title}{Space object collision probability using
  multidirectional gaussian mixture models},
\newblock \bibinfo{journal}{Journal of Guidance, Control, and Dynamics}
  \bibinfo{volume}{39} (\bibinfo{year}{2016}) \bibinfo{pages}{2163--2169}.
%Type = Inproceedings
\bibitem[{Horwood and Poore(2012)}]{Horwood2012}
\bibinfo{author}{J.~T. Horwood}, \bibinfo{author}{A.~B. Poore},
\newblock \bibinfo{title}{{Orbital State Uncertainty Realism}},
\newblock in: \bibinfo{booktitle}{Advanced Maui Optical and Space Surveillance
  Technologies Conference}, \bibinfo{address}{Wailea, HI},
  \bibinfo{year}{2012}, p.~\bibinfo{pages}{48}.
%Type = Inproceedings
\bibitem[{Horwood et~al.(2014)Horwood, Aristoff, Singh, and
  Poore}]{Horwood2014}
\bibinfo{author}{J.~T. Horwood}, \bibinfo{author}{J.~M. Aristoff},
  \bibinfo{author}{N.~Singh}, \bibinfo{author}{A.~B. Poore},
\newblock \bibinfo{title}{{A comparative study of new non-linear uncertainty
  propagation methods for space surveillance}},
\newblock in: \bibinfo{booktitle}{Signal and Data Processing of Small Targets
  2014}, volume \bibinfo{volume}{9092}, \bibinfo{address}{Baltimore, MD},
  \bibinfo{year}{2014}, p.~\bibinfo{pages}{12}.
%Type = Article
\bibitem[{Horwood and Poore(2014)}]{horwood2014a}
\bibinfo{author}{J.~T. Horwood}, \bibinfo{author}{A.~B. Poore},
\newblock \bibinfo{title}{Gauss von mises distribution for improved uncertainty
  realism in space situational awareness},
\newblock \bibinfo{journal}{SIAM/ASA Journal on Uncertainty Quantification}
  \bibinfo{volume}{2} (\bibinfo{year}{2014}) \bibinfo{pages}{276--304}.
%Type = Book
\bibitem[{Risken(1989)}]{Risken_1989}
\bibinfo{author}{H.~Risken}, \bibinfo{title}{The Fokker-Planck Equation},
  Springer Series in Synergetics, \bibinfo{publisher}{Springer Berlin
  Heidelberg}, \bibinfo{address}{Berlin, Germany}, \bibinfo{year}{1989}.
%Type = Incollection
\bibitem[{Jazwinski(1970)}]{jazwinski2007stochastic}
\bibinfo{author}{A.~H. Jazwinski},
\newblock \bibinfo{title}{Stochastic differential equations},
\newblock in: \bibinfo{booktitle}{Stochastic Processes and Filtering Theory},
  volume~\bibinfo{volume}{64} of \textit{\bibinfo{series}{Mathematics in
  Science and Engineering}}, \bibinfo{publisher}{Elsevier},
  \bibinfo{year}{1970}, pp. \bibinfo{pages}{93--141}.
%Type = Inproceedings
\bibitem[{Jones and Anderson(2012)}]{Jones2012}
\bibinfo{author}{B.~Jones}, \bibinfo{author}{R.~Anderson},
\newblock \bibinfo{title}{A survey of symplectic and collocation methods for
  orbit propagation},
\newblock in: \bibinfo{booktitle}{22nd AAS/AIAA Space Flight Mechanics
  Meeting}, \bibinfo{address}{Charleston, SC}, \bibinfo{year}{2012},
  p.~\bibinfo{pages}{20}.
%Type = Inproceedings
\bibitem[{Bradley et~al.(2012)Bradley, Jones, Beylkin, and
  Axelrad}]{Bradley2012}
\bibinfo{author}{B.~Bradley}, \bibinfo{author}{B.~Jones},
  \bibinfo{author}{G.~Beylkin}, \bibinfo{author}{P.~Axelrad},
\newblock \bibinfo{title}{A new numerical integration technique in
  astrodynamics},
\newblock in: \bibinfo{booktitle}{Proceedings of the 22nd Annual AAS/AIAA
  Spaceflight Mechanics Meeting}, \bibinfo{address}{Charleston, SC},
  \bibinfo{year}{2012}, pp. \bibinfo{pages}{1--20}.
%Type = Article
\bibitem[{Bai and Junkins(2011)}]{Bai2011}
\bibinfo{author}{X.~Bai}, \bibinfo{author}{J.~L. Junkins},
\newblock \bibinfo{title}{Modified chebyshev-picard iteration methods for orbit
  propagation},
\newblock \bibinfo{journal}{The Journal of the Astronautical Sciences}
  \bibinfo{volume}{58} (\bibinfo{year}{2011}) \bibinfo{pages}{583--613}.
%Type = Inbook
\bibitem[{Butcher(2016)}]{Butcher}
\bibinfo{author}{J.~C. Butcher}, \bibinfo{title}{Runge–Kutta Methods},
  \bibinfo{edition}{3} ed., \bibinfo{publisher}{John Wiley \& Sons, Ltd},
  \bibinfo{address}{West Sussex, England}, \bibinfo{year}{2016}, pp.
  \bibinfo{pages}{143--331}.
%Type = Article
\bibitem[{Aristoff et~al.(2014)Aristoff, Horwood, and Poore}]{Aristoff2014}
\bibinfo{author}{J.~M. Aristoff}, \bibinfo{author}{J.~T. Horwood},
  \bibinfo{author}{A.~B. Poore},
\newblock \bibinfo{title}{{Orbit and uncertainty propagation: A comparison of
  Gauss-Legendre-, Dormand-Prince-, and Chebyshev-Picard-based approaches}},
\newblock \bibinfo{journal}{Celestial Mechanics and Dynamical Astronomy}
  \bibinfo{volume}{118} (\bibinfo{year}{2014}) \bibinfo{pages}{13--28}.
%Type = Article
\bibitem[{Sharma and James~Raj(1988)}]{Sharma1988}
\bibinfo{author}{R.~K. Sharma}, \bibinfo{author}{M.~X. James~Raj},
\newblock \bibinfo{title}{Long-term orbit computations with ks uniformly
  regular canonical elements with oblateness},
\newblock \bibinfo{journal}{Earth, Moon, and Planets} \bibinfo{volume}{42}
  (\bibinfo{year}{1988}) \bibinfo{pages}{163--178}.
%Type = Article
\bibitem[{Kustaanheimo et~al.(1965)Kustaanheimo, Schinzel, Davenport, and
  Stiefel}]{Kustaanheimo}
\bibinfo{author}{P.~Kustaanheimo}, \bibinfo{author}{A.~Schinzel},
  \bibinfo{author}{H.~Davenport}, \bibinfo{author}{E.~Stiefel},
\newblock \bibinfo{title}{Perturbation theory of kepler motion based on spinor
  regularization.},
\newblock \bibinfo{journal}{Journal für die reine und angewandte Mathematik}
  \bibinfo{volume}{1965} (\bibinfo{year}{1965}) \bibinfo{pages}{204--219}.
%Type = Article
\bibitem[{Sperling(1961)}]{sperling1961computation}
\bibinfo{author}{H.~Sperling},
\newblock \bibinfo{title}{Computation of keplerian conic sections},
\newblock \bibinfo{journal}{American Rocket Society journal}
  \bibinfo{volume}{31} (\bibinfo{year}{1961}) \bibinfo{pages}{660--661}.
%Type = Article
\bibitem[{Burdet(1967)}]{burdet1967regularization}
\bibinfo{author}{C.~A. Burdet},
\newblock \bibinfo{title}{Regularization of the two body problem},
\newblock \bibinfo{journal}{Zeitschrift f{\"u}r angewandte Mathematik und
  Physik ZAMP} \bibinfo{volume}{18} (\bibinfo{year}{1967})
  \bibinfo{pages}{434--438}.
%Type = Article
\bibitem[{Baù et~al.(2014)Baù, Urrutxua, and Pelaez}]{bau2014}
\bibinfo{author}{G.~Baù}, \bibinfo{author}{H.~Urrutxua},
  \bibinfo{author}{J.~Pelaez},
\newblock \bibinfo{title}{Edromo: An accurate propagator for elliptical orbits
  in the perturbed two-body problem},
\newblock \bibinfo{journal}{Advances in the Astronautical Sciences}
  \bibinfo{volume}{152} (\bibinfo{year}{2014}) \bibinfo{pages}{379--399}.
%Type = Article
\bibitem[{Urrutxua et~al.(2015)Urrutxua, Sanjurjo-Rivo, and
  Pel{\'{a}}ez}]{Urrutxua_2015}
\bibinfo{author}{H.~Urrutxua}, \bibinfo{author}{M.~Sanjurjo-Rivo},
  \bibinfo{author}{J.~Pel{\'{a}}ez},
\newblock \bibinfo{title}{{DROMO} propagator revisited},
\newblock \bibinfo{journal}{Celestial Mechanics and Dynamical Astronomy}
  \bibinfo{volume}{124} (\bibinfo{year}{2015}) \bibinfo{pages}{1--31}.
%Type = Phdthesis
\bibitem[{Zwiep(2009)}]{Zwiep2009}
\bibinfo{author}{M.~N. Zwiep}, \bibinfo{title}{{Comparison and Design of
  Simplified General Perturbation Models (SGP4) and Code for NASA Johnson Space
  Center, Orbital Debris Program Office}}, Ph.D. thesis, California Polytechnic
  State University, \bibinfo{address}{San Luis Obispo, CA},
  \bibinfo{year}{2009}. \bibinfo{note}{{Digital Commons Cal Poly}}.
%Type = Inproceedings
\bibitem[{Vallado et~al.(2013)Vallado, Virgili, and Flohrer}]{Vallado2013}
\bibinfo{author}{D.~A. Vallado}, \bibinfo{author}{B.~B. Virgili},
  \bibinfo{author}{T.~Flohrer},
\newblock \bibinfo{title}{{Improved SSA Through Orbit Determination of Two-Line
  Element Sets}},
\newblock in: \bibinfo{booktitle}{6th European Conference on Space Debris},
  volume~\bibinfo{volume}{6}, \bibinfo{publisher}{ESA}, \bibinfo{year}{2013},
  pp. \bibinfo{pages}{22--25}.
%Type = Incollection
\bibitem[{Chen et~al.(2017)Chen, Bai, Liang, and Li}]{Lei2017}
\bibinfo{author}{L.~Chen}, \bibinfo{author}{X.-Z. Bai}, \bibinfo{author}{Y.-G.
  Liang}, \bibinfo{author}{K.-B. Li},
\newblock \bibinfo{title}{Orbital prediction error propagation of space
  objects},
\newblock in: \bibinfo{booktitle}{Orbital Data Applications for Space Objects:
  Conjunction Assessment and Situation Analysis}, \bibinfo{edition}{1st} ed.,
  \bibinfo{publisher}{Springer}, \bibinfo{address}{Singapore},
  \bibinfo{year}{2017}, pp. \bibinfo{pages}{23--75}.
%Type = Article
\bibitem[{Levit and Marshall(2011)}]{Levit2011}
\bibinfo{author}{C.~Levit}, \bibinfo{author}{W.~Marshall},
\newblock \bibinfo{title}{{Improved Orbit Predictions Using Two-Line
  Elements}},
\newblock \bibinfo{journal}{Advances in Space Research} \bibinfo{volume}{47}
  (\bibinfo{year}{2011}) \bibinfo{pages}{1107--1115}.
%Type = Inproceedings
\bibitem[{Bennett et~al.(2012)Bennett, Sang, Smith, and Zhang}]{Bennett2012}
\bibinfo{author}{J.~C. Bennett}, \bibinfo{author}{J.~Sang},
  \bibinfo{author}{C.~H. Smith}, \bibinfo{author}{K.~Zhang},
\newblock \bibinfo{title}{{Improving Low-Earth Orbit Predictions Using Two-line
  Element Data with Bias Correction}},
\newblock in: \bibinfo{booktitle}{Advanced Maui Optical and Space Surveillance
  Technologies Conference}, \bibinfo{address}{Wailea, HI},
  \bibinfo{year}{2012}, p.~\bibinfo{pages}{46}.
%Type = Article
\bibitem[{Sang et~al.(2017)Sang, Li, Chen, Zhang, and Ning}]{Sang2017}
\bibinfo{author}{J.~Sang}, \bibinfo{author}{B.~Li}, \bibinfo{author}{J.~Chen},
  \bibinfo{author}{P.~Zhang}, \bibinfo{author}{J.~Ning},
\newblock \bibinfo{title}{{Analytical representations of precise orbit
  predictions for Earth orbiting space objects}},
\newblock \bibinfo{journal}{Advances in Space Research} \bibinfo{volume}{59}
  (\bibinfo{year}{2017}) \bibinfo{pages}{698--714}.
%Type = Article
\bibitem[{San-Juan et~al.(2017)San-Juan, Pérez, San-Martín, and
  Vergara}]{SANJUAN2017254}
\bibinfo{author}{J.~F. San-Juan}, \bibinfo{author}{I.~Pérez},
  \bibinfo{author}{M.~San-Martín}, \bibinfo{author}{E.~P. Vergara},
\newblock \bibinfo{title}{Hybrid {SGP4} orbit propagator},
\newblock \bibinfo{journal}{Acta Astronautica} \bibinfo{volume}{137}
  (\bibinfo{year}{2017}) \bibinfo{pages}{254--260}.
%Type = Article
\bibitem[{Peng and Bai(2020)}]{Peng2020}
\bibinfo{author}{H.~Peng}, \bibinfo{author}{X.~Bai},
\newblock \bibinfo{title}{{Machine Learning Approach to Improve Satellite Orbit
  Prediction Accuracy Using Publicly Available Data}},
\newblock \bibinfo{journal}{Journal of the Astronautical Sciences}
  \bibinfo{volume}{67} (\bibinfo{year}{2020}) \bibinfo{pages}{762--793}.
%Type = Inproceedings
\bibitem[{Muldoon et~al.(2009)Muldoon, Elkaim, Rickard, and
  Weeden}]{Muldoon2009}
\bibinfo{author}{A.~R. Muldoon}, \bibinfo{author}{G.~H. Elkaim},
  \bibinfo{author}{I.~F. Rickard}, \bibinfo{author}{B.~Weeden},
\newblock \bibinfo{title}{{Improved Orbital Debris Trajectory Estimation Based
  on Sequential TLE Processing}},
\newblock in: \bibinfo{booktitle}{60th International Astronautical Congress
  2009}, volume~\bibinfo{volume}{3}, \bibinfo{publisher}{International
  Astronautical Federation (IAF)}, \bibinfo{address}{Daejeon, South Korea},
  \bibinfo{year}{2009}, pp. \bibinfo{pages}{1864--1869}.
%Type = Inproceedings
\bibitem[{Rautalin et~al.(2018)Rautalin, Ali-L{\"{o}}ytty, and
  Pich{\'{e}}}]{Rautalin2018}
\bibinfo{author}{S.~Rautalin}, \bibinfo{author}{S.~Ali-L{\"{o}}ytty},
  \bibinfo{author}{R.~Pich{\'{e}}},
\newblock \bibinfo{title}{{Latent force models in autonomous GNSS satellite
  orbit prediction}},
\newblock in: \bibinfo{booktitle}{2017 International Conference on Localization
  and GNSS, ICL-GNSS 2017}, \bibinfo{publisher}{Institute of Electric and
  Electronics Engineers (IEEE)}, \bibinfo{address}{Piscataway, New Jersey},
  \bibinfo{year}{2018}, pp. \bibinfo{pages}{1--6}.
%Type = Article
\bibitem[{Peng and Bai(2018)}]{Peng2018a}
\bibinfo{author}{H.~Peng}, \bibinfo{author}{X.~Bai},
\newblock \bibinfo{title}{{Exploring Capability of Support Vector Machine for
  Improving Satellite Orbit Prediction Accuracy}},
\newblock \bibinfo{journal}{Journal of Aerospace Information Systems}
  \bibinfo{volume}{15} (\bibinfo{year}{2018}) \bibinfo{pages}{366--381}.
%Type = Article
\bibitem[{Peng and Bai(2019{\natexlab{a}})}]{Peng2019}
\bibinfo{author}{H.~Peng}, \bibinfo{author}{X.~Bai},
\newblock \bibinfo{title}{{Comparative Evaluation of Three Machine Learning
  Algorithms on Improving Orbit Prediction Accuracy}},
\newblock \bibinfo{journal}{Astrodynamics} \bibinfo{volume}{3}
  (\bibinfo{year}{2019}{\natexlab{a}}) \bibinfo{pages}{325--343}.
%Type = Article
\bibitem[{Peng and Bai(2019{\natexlab{b}})}]{Peng2019a}
\bibinfo{author}{H.~Peng}, \bibinfo{author}{X.~Bai},
\newblock \bibinfo{title}{{Gaussian Processes for improving orbit prediction
  accuracy}},
\newblock \bibinfo{journal}{Acta Astronautica} \bibinfo{volume}{161}
  (\bibinfo{year}{2019}{\natexlab{b}}) \bibinfo{pages}{44--56}.
%Type = Article
\bibitem[{Peng and Bai(2021)}]{Peng2021}
\bibinfo{author}{H.~Peng}, \bibinfo{author}{X.~Bai},
\newblock \bibinfo{title}{{Fusion of a machine learning approach and classical
  orbit predictions}},
\newblock \bibinfo{journal}{Acta Astronautica} \bibinfo{volume}{184}
  (\bibinfo{year}{2021}) \bibinfo{pages}{222--240}.
%Type = Article
\bibitem[{Li et~al.(2021)Li, Zhang, Huang, and Sang}]{Li2021}
\bibinfo{author}{B.~Li}, \bibinfo{author}{Y.~Zhang},
  \bibinfo{author}{J.~Huang}, \bibinfo{author}{J.~Sang},
\newblock \bibinfo{title}{{Improved orbit predictions using two-line elements
  through error pattern mining and transferring}},
\newblock \bibinfo{journal}{Acta Astronautica} \bibinfo{volume}{188}
  (\bibinfo{year}{2021}) \bibinfo{pages}{405--415}.
%Type = Inproceedings
\bibitem[{Pihlajasalo et~al.(2018)Pihlajasalo, Lepp{\"{a}}koski,
  Ali-L{\"{o}}ytty, and Pich{\'{e}}}]{Pihlajasalo2018}
\bibinfo{author}{J.~Pihlajasalo}, \bibinfo{author}{H.~Lepp{\"{a}}koski},
  \bibinfo{author}{S.~Ali-L{\"{o}}ytty}, \bibinfo{author}{R.~Pich{\'{e}}},
\newblock \bibinfo{title}{{Improvement of GPS and BeiDou extended orbit
  predictions with CNNs}},
\newblock in: \bibinfo{booktitle}{2018 European Navigation Conference, ENC
  2018}, \bibinfo{publisher}{Institute of Electric and Electronics Engineers
  (IEEE)}, \bibinfo{address}{Gothenburg, Sweden}, \bibinfo{year}{2018}, pp.
  \bibinfo{pages}{54--59}.
%Type = Inproceedings
\bibitem[{San-Juan et~al.(2018)San-Juan, P{\'{e}}rez, Vergara, {Montserrat
  San}, L{\'{o}}pez, Wittig, and Izzo}]{San-Juan2018}
\bibinfo{author}{J.~F. San-Juan}, \bibinfo{author}{I.~P{\'{e}}rez},
  \bibinfo{author}{E.~Vergara}, \bibinfo{author}{M.~{Montserrat San}},
  \bibinfo{author}{R.~L{\'{o}}pez}, \bibinfo{author}{A.~Wittig},
  \bibinfo{author}{D.~Izzo},
\newblock \bibinfo{title}{{Hybrid SGP4 propagator based on machine-learning
  techniques applied to GALILEO-type orbits}},
\newblock in: \bibinfo{booktitle}{69th International Astronautical Congress
  (IAC-18)}, \bibinfo{publisher}{International Astronautical Federation (IAF)},
  \bibinfo{address}{Bremen, Germany}, \bibinfo{year}{2018}, pp.
  \bibinfo{pages}{1--5}.
%Type = Article
\bibitem[{Curzi et~al.(2022)Curzi, Modenini, and Tortora}]{Curzi2022}
\bibinfo{author}{G.~Curzi}, \bibinfo{author}{D.~Modenini},
  \bibinfo{author}{P.~Tortora},
\newblock \bibinfo{title}{{Two-line-element propagation improvement and
  uncertainty estimation using recurrent neural networks}},
\newblock \bibinfo{journal}{CEAS Space Journal} \bibinfo{volume}{14}
  (\bibinfo{year}{2022}) \bibinfo{pages}{197--204}.
%Type = Inproceedings
\bibitem[{Salleh et~al.(2021)Salleh, Mohd~Azmi, and Yuhaniz}]{Salleh2021}
\bibinfo{author}{N.~Salleh}, \bibinfo{author}{N.~F. Mohd~Azmi},
  \bibinfo{author}{S.~S. Yuhaniz},
\newblock \bibinfo{title}{An adaptation of deep learning technique in orbit
  propagation model using long short-term memory},
\newblock in: \bibinfo{booktitle}{2021 International Conference on Electrical,
  Communication, and Computer Engineering (ICECCE)},
  \bibinfo{publisher}{Institute of Electric and Electronics Engineers (IEEE)},
  \bibinfo{address}{Kuala Lumpur, Malaysia}, \bibinfo{year}{2021}, pp.
  \bibinfo{pages}{1--6}.
%Type = Article
\bibitem[{Li et~al.(2020)Li, Huang, Feng, Wang, and Sang}]{Li2020}
\bibinfo{author}{B.~Li}, \bibinfo{author}{J.~Huang}, \bibinfo{author}{Y.~Feng},
  \bibinfo{author}{F.~Wang}, \bibinfo{author}{J.~Sang},
\newblock \bibinfo{title}{{A Machine Learning-Based Approach for Improved Orbit
  Predictions of LEO Space Debris with Sparse Tracking Data from a Single
  Station}},
\newblock \bibinfo{journal}{Institute of Electric and Electronics Engineers
  (IEEE) Transactions on Aerospace and Electronic Systems} \bibinfo{volume}{56}
  (\bibinfo{year}{2020}) \bibinfo{pages}{4253--4268}.
%Type = Article
\bibitem[{Peng and Bai(2018)}]{Peng2018b}
\bibinfo{author}{H.~Peng}, \bibinfo{author}{X.~Bai},
\newblock \bibinfo{title}{{Improving orbit prediction accuracy through
  supervised machine learning}},
\newblock \bibinfo{journal}{Advances in Space Research} \bibinfo{volume}{61}
  (\bibinfo{year}{2018}) \bibinfo{pages}{2628--2646}.
%Type = Article
\bibitem[{Emmert(2015)}]{Emmert2015}
\bibinfo{author}{J.~T. Emmert},
\newblock \bibinfo{title}{{Thermospheric mass density: a review}},
\newblock \bibinfo{journal}{Advances in Space Research} \bibinfo{volume}{56}
  (\bibinfo{year}{2015}) \bibinfo{pages}{773--824}.
%Type = Incollection
\bibitem[{Bowman et~al.(2008)Bowman, Tobiska, Marcos, Huang, Lin, and
  Burke}]{Bowman2008}
\bibinfo{author}{B.~R. Bowman}, \bibinfo{author}{W.~K. Tobiska},
  \bibinfo{author}{F.~A. Marcos}, \bibinfo{author}{C.~Y. Huang},
  \bibinfo{author}{C.~S. Lin}, \bibinfo{author}{W.~J. Burke},
\newblock \bibinfo{title}{{A new empirical thermospheric density model JB2008
  using new solar and geomagnetic indices}},
\newblock in: \bibinfo{booktitle}{AIAA/AAS Astrodynamics Specialist Conference
  and Exhibit}, \bibinfo{publisher}{AIAA}, \bibinfo{address}{Honolulu, HI},
  \bibinfo{year}{2008}, p.~\bibinfo{pages}{19}.
%Type = Article
\bibitem[{Bruinsma and Boniface(2021)}]{Bruinsma2021}
\bibinfo{author}{S.~Bruinsma}, \bibinfo{author}{C.~Boniface},
\newblock \bibinfo{title}{{The operational and research DTM-2020 thermosphere
  models}},
\newblock \bibinfo{journal}{Journal of Space Weather and Space Climate}
  \bibinfo{volume}{11} (\bibinfo{year}{2021}) \bibinfo{pages}{15}.
%Type = Article
\bibitem[{Emmert et~al.(2021)Emmert, Drob, Picone, Siskind, Jones, Mlynczak,
  Bernath, Chu, Doornbos, Funke, Goncharenko, Hervig, Schwartz, Sheese, Vargas,
  Williams, and Yuan}]{Emmert2021}
\bibinfo{author}{J.~T. Emmert}, \bibinfo{author}{D.~P. Drob},
  \bibinfo{author}{J.~M. Picone}, \bibinfo{author}{D.~E. Siskind},
  \bibinfo{author}{M.~Jones}, \bibinfo{author}{M.~G. Mlynczak},
  \bibinfo{author}{P.~F. Bernath}, \bibinfo{author}{X.~Chu},
  \bibinfo{author}{E.~Doornbos}, \bibinfo{author}{B.~Funke},
  \bibinfo{author}{L.~P. Goncharenko}, \bibinfo{author}{M.~E. Hervig},
  \bibinfo{author}{M.~J. Schwartz}, \bibinfo{author}{P.~E. Sheese},
  \bibinfo{author}{F.~Vargas}, \bibinfo{author}{B.~P. Williams},
  \bibinfo{author}{T.~Yuan},
\newblock \bibinfo{title}{{NRLMSIS} 2.0: A whole-atmosphere empirical model of
  temperature and neutral species densities},
\newblock \bibinfo{journal}{Earth and Space Science} \bibinfo{volume}{8}
  (\bibinfo{year}{2021}).
%Type = Article
\bibitem[{Hedin(1979)}]{HEDIN1}
\bibinfo{author}{A.~E. Hedin},
\newblock \bibinfo{title}{Neutral thermospheric composition and thermal
  structure},
\newblock \bibinfo{journal}{Reviews of Geophysics} \bibinfo{volume}{17}
  (\bibinfo{year}{1979}) \bibinfo{pages}{477--485}.
%Type = Article
\bibitem[{Hedin(1983)}]{hedin2}
\bibinfo{author}{A.~E. Hedin},
\newblock \bibinfo{title}{A revised thermospheric model based on mass
  spectrometer and incoherent scatter data: {MSIS-83}},
\newblock \bibinfo{journal}{Journal of Geophysical Research: Space Physics}
  \bibinfo{volume}{88} (\bibinfo{year}{1983}) \bibinfo{pages}{10170--10188}.
%Type = Article
\bibitem[{Hedin(1987)}]{hedin3}
\bibinfo{author}{A.~E. Hedin},
\newblock \bibinfo{title}{{MSIS-86 Thermospheric Model}},
\newblock \bibinfo{journal}{Journal of Geophysical Research: Space Physics}
  \bibinfo{volume}{92} (\bibinfo{year}{1987}) \bibinfo{pages}{4649--4662}.
%Type = Article
\bibitem[{Hedin(1991)}]{hedin4}
\bibinfo{author}{A.~E. Hedin},
\newblock \bibinfo{title}{Extension of the msis thermosphere model into the
  middle and lower atmosphere},
\newblock \bibinfo{journal}{Journal of Geophysical Research: Space Physics}
  \bibinfo{volume}{96} (\bibinfo{year}{1991}) \bibinfo{pages}{1159--1172}.
%Type = Article
\bibitem[{Picone et~al.(2002)Picone, Hedin, Drob, and Aikin}]{picone}
\bibinfo{author}{J.~M. Picone}, \bibinfo{author}{A.~E. Hedin},
  \bibinfo{author}{D.~P. Drob}, \bibinfo{author}{A.~C. Aikin},
\newblock \bibinfo{title}{{NRLMSISE-00 empirical model of the atmosphere:
  Statistical comparisons and scientific issues}},
\newblock \bibinfo{journal}{Journal of Geophysical Research: Space Physics}
  \bibinfo{volume}{107} (\bibinfo{year}{2002}).
%Type = Article
\bibitem[{{Barlier} et~al.(1978){Barlier}, {Berger}, {Falin}, {Kockarts}, and
  {Thuillier}}]{barlier78}
\bibinfo{author}{F.~{Barlier}}, \bibinfo{author}{C.~{Berger}},
  \bibinfo{author}{J.~L. {Falin}}, \bibinfo{author}{G.~{Kockarts}},
  \bibinfo{author}{G.~{Thuillier}},
\newblock \bibinfo{title}{{A thermospheric model based on satellite drag
  data.}},
\newblock \bibinfo{journal}{Annales de Geophysique} \bibinfo{volume}{34}
  (\bibinfo{year}{1978}) \bibinfo{pages}{9--24}.
%Type = Article
\bibitem[{Berger et~al.(1998)Berger, Biancale, Ill, and Barlier}]{Berger1998}
\bibinfo{author}{C.~Berger}, \bibinfo{author}{R.~Biancale},
  \bibinfo{author}{M.~Ill}, \bibinfo{author}{F.~Barlier},
\newblock \bibinfo{title}{Improvement of the empirical thermospheric model
  {DTM}: {DTM}94 -- a comparative review of various temporal variations and
  prospects in space geodesy applications},
\newblock \bibinfo{journal}{Journal of Geodesy} \bibinfo{volume}{72}
  (\bibinfo{year}{1998}) \bibinfo{pages}{161--178}.
%Type = Article
\bibitem[{Bruinsma and Biancale(2003)}]{BRUINSMA2003}
\bibinfo{author}{S.~Bruinsma}, \bibinfo{author}{R.~Biancale},
\newblock \bibinfo{title}{Total densities derived from accelerometer data},
\newblock \bibinfo{journal}{Journal of Spacecraft and Rockets}
  \bibinfo{volume}{40} (\bibinfo{year}{2003}) \bibinfo{pages}{230--236}.
%Type = Article
\bibitem[{Bruinsma et~al.(2012)Bruinsma, Sánchez-Ortiz, Olmedo, and
  Guijarro}]{bruinsma2012}
\bibinfo{author}{S.~Bruinsma}, \bibinfo{author}{N.~Sánchez-Ortiz},
  \bibinfo{author}{E.~Olmedo}, \bibinfo{author}{N.~Guijarro},
\newblock \bibinfo{title}{{Evaluation of the DTM-2009 thermosphere model for
  benchmarking purposes}},
\newblock \bibinfo{journal}{Journal of Space Weather and Space Climate}
  \bibinfo{volume}{2} (\bibinfo{year}{2012}) \bibinfo{pages}{4}.
%Type = Article
\bibitem[{Jacchia(1961)}]{JACCHIA1961}
\bibinfo{author}{L.~G. Jacchia},
\newblock \bibinfo{title}{A working model for the upper atmosphere},
\newblock \bibinfo{journal}{Nature} \bibinfo{volume}{192}
  (\bibinfo{year}{1961}) \bibinfo{pages}{1147--1148}.
%Type = Article
\bibitem[{{Jacchia}(1965)}]{Jacchia65}
\bibinfo{author}{L.~G. {Jacchia}},
\newblock \bibinfo{title}{Static diffusion models of the upper atmosphere with
  empirical temperature profiles},
\newblock \bibinfo{journal}{Smithsonian Contributions to Astrophysics}
  \bibinfo{volume}{8} (\bibinfo{year}{1965}) \bibinfo{pages}{213–257}.
%Type = Article
\bibitem[{{Jacchia}(1970)}]{j70}
\bibinfo{author}{L.~G. {Jacchia}},
\newblock \bibinfo{title}{{New Static Models of the Thermosphere and Exosphere
  with Empirical Temperature Profiles}},
\newblock \bibinfo{journal}{SAO Special Report} \bibinfo{volume}{313}
  (\bibinfo{year}{1970}).
%Type = Article
\bibitem[{{Jacchia}(1971)}]{71J}
\bibinfo{author}{L.~G. {Jacchia}},
\newblock \bibinfo{title}{{Revised Static Models of the Thermosphere and
  Exosphere with Empirical Temperature Profiles}},
\newblock \bibinfo{journal}{SAO Special Report} \bibinfo{volume}{332}
  (\bibinfo{year}{1971}).
%Type = Techreport
\bibitem[{Suggs and Suggs(2017)}]{Suggs2017MarshallET}
\bibinfo{author}{R.~J. Suggs}, \bibinfo{author}{R.~M. Suggs},
  \bibinfo{title}{Marshall Engineering Termosphere Model, Version MET-2007},
  \bibinfo{type}{Technical Report} \bibinfo{number}{NASA/TM—2017–218238},
  Marshall Space Flight Center, Huntsville, Alabama, \bibinfo{year}{2017}.
%Type = Article
\bibitem[{Justus et~al.(2004)Justus, Duvall, and Johnson}]{JUSTUS20041731}
\bibinfo{author}{C.~Justus}, \bibinfo{author}{A.~Duvall},
  \bibinfo{author}{D.~Johnson},
\newblock \bibinfo{title}{Earth global reference atmospheric model (gram-99)
  and trace constituents},
\newblock \bibinfo{journal}{Advances in Space Research} \bibinfo{volume}{34}
  (\bibinfo{year}{2004}) \bibinfo{pages}{1731--1735}.
%Type = Article
\bibitem[{Bowman et~al.(2008)Bowman, {Kent Tobiska}, Marcos, and
  Valladares}]{BOWMAN2008774}
\bibinfo{author}{B.~R. Bowman}, \bibinfo{author}{W.~{Kent Tobiska}},
  \bibinfo{author}{F.~A. Marcos}, \bibinfo{author}{C.~Valladares},
\newblock \bibinfo{title}{{The JB2006 empirical thermospheric density model}},
\newblock \bibinfo{journal}{Journal of Atmospheric and Solar-Terrestrial
  Physics} \bibinfo{volume}{70} (\bibinfo{year}{2008})
  \bibinfo{pages}{774--793}.
%Type = Article
\bibitem[{Vallado and Finkleman(2014)}]{Vallado2014a}
\bibinfo{author}{D.~A. Vallado}, \bibinfo{author}{D.~Finkleman},
\newblock \bibinfo{title}{{A critical assessment of satellite drag and
  atmospheric density modeling}},
\newblock \bibinfo{journal}{Acta Astronautica} \bibinfo{volume}{95}
  (\bibinfo{year}{2014}) \bibinfo{pages}{141--165}.
%Type = Article
\bibitem[{He et~al.(2018)He, Yang, Carter, Kerr, Wu, Deleflie, Cai, Zhang,
  Sagnières, and Norman}]{HE201831}
\bibinfo{author}{C.~He}, \bibinfo{author}{Y.~Yang},
  \bibinfo{author}{B.~Carter}, \bibinfo{author}{E.~Kerr},
  \bibinfo{author}{S.~Wu}, \bibinfo{author}{F.~Deleflie},
  \bibinfo{author}{H.~Cai}, \bibinfo{author}{K.~Zhang},
  \bibinfo{author}{L.~Sagnières}, \bibinfo{author}{R.~Norman},
\newblock \bibinfo{title}{Review and comparison of empirical thermospheric mass
  density models},
\newblock \bibinfo{journal}{Progress in Aerospace Sciences}
  \bibinfo{volume}{103} (\bibinfo{year}{2018}) \bibinfo{pages}{31--51}.
%Type = Inproceedings
\bibitem[{Emmert et~al.(2014)Emmert, Byers, Warren, and Segerman}]{Emmert2014}
\bibinfo{author}{J.~Emmert}, \bibinfo{author}{J.~Byers},
  \bibinfo{author}{H.~Warren}, \bibinfo{author}{A.~Segerman},
\newblock \bibinfo{title}{{Propagation of Forecast Errors from the Sun to LEO
  Trajectories: How Does Drag Uncertainty Affect Conjunction Frequency?}},
\newblock in: \bibinfo{booktitle}{Advanced Maui Optical and Space Surveillance
  Technologies Conference}, \bibinfo{address}{Wailea, HI},
  \bibinfo{year}{2014}, p.~\bibinfo{pages}{8}.
%Type = Article
\bibitem[{Williams(1991)}]{Williams1991}
\bibinfo{author}{K.~E. Williams},
\newblock \bibinfo{title}{{Prediction of solar activity with a neural network
  and its effect on orbit prediction}},
\newblock \bibinfo{journal}{Johns Hopkins APL Technical Digest (Applied Physics
  Laboratory)} \bibinfo{volume}{12} (\bibinfo{year}{1991})
  \bibinfo{pages}{310--317}.
%Type = Article
\bibitem[{Gleisner et~al.(1996)Gleisner, Lundstedt, and
  Wintoft}]{Gleisner1996PredictingGS}
\bibinfo{author}{H.~Gleisner}, \bibinfo{author}{H.~Lundstedt},
  \bibinfo{author}{P.~Wintoft},
\newblock \bibinfo{title}{Predicting geomagnetic storms from solar-wind data
  using time-delay neural networks},
\newblock \bibinfo{journal}{Annales Geophysicae} \bibinfo{volume}{14}
  (\bibinfo{year}{1996}) \bibinfo{pages}{679--686}.
%Type = Article
\bibitem[{Huang et~al.(2009)Huang, Liu, and Wang}]{Huang2009}
\bibinfo{author}{C.~Huang}, \bibinfo{author}{D.-D. Liu}, \bibinfo{author}{J.-S.
  Wang},
\newblock \bibinfo{title}{Forecast daily indices of solar activity, {F10. 7},
  using support vector regression method},
\newblock \bibinfo{journal}{Research in Astronomy and Astrophysics}
  \bibinfo{volume}{9} (\bibinfo{year}{2009}) \bibinfo{pages}{694--702}.
%Type = Article
\bibitem[{{Yaya} et~al.(2017){Yaya}, {Hecker}, {Dudok de Wit}, {F{\`e}vre}, and
  {Bruinsma}}]{Yaya2017}
\bibinfo{author}{P.~{Yaya}}, \bibinfo{author}{L.~{Hecker}},
  \bibinfo{author}{T.~{Dudok de Wit}}, \bibinfo{author}{C.~L. {F{\`e}vre}},
  \bibinfo{author}{S.~{Bruinsma}},
\newblock \bibinfo{title}{{Solar radio proxies for improved satellite orbit
  prediction}},
\newblock \bibinfo{journal}{Journal of Space Weather and Space Climate}
  \bibinfo{volume}{7} (\bibinfo{year}{2017}) \bibinfo{pages}{17}.
%Type = Article
\bibitem[{Tobiska et~al.(2000)Tobiska, Woods, Eparvier, Viereck, Floyd, Bouwer,
  Rottman, and White}]{Tobiska2000}
\bibinfo{author}{W.~K. Tobiska}, \bibinfo{author}{T.~Woods},
  \bibinfo{author}{F.~Eparvier}, \bibinfo{author}{R.~Viereck},
  \bibinfo{author}{L.~Floyd}, \bibinfo{author}{D.~Bouwer},
  \bibinfo{author}{G.~Rottman}, \bibinfo{author}{O.~White},
\newblock \bibinfo{title}{{The SOLAR2000 empirical solar irradiance model and
  forecast tool}},
\newblock \bibinfo{journal}{Journal of Atmospheric and Solar-Terrestrial
  Physics} \bibinfo{volume}{62} (\bibinfo{year}{2000})
  \bibinfo{pages}{1233--1250}.
%Type = Article
\bibitem[{Warren et~al.(2017)Warren, Emmert, and Crump}]{warren2017}
\bibinfo{author}{H.~P. Warren}, \bibinfo{author}{J.~T. Emmert},
  \bibinfo{author}{N.~A. Crump},
\newblock \bibinfo{title}{Linear forecasting of the {F10.7} proxy for solar
  activity},
\newblock \bibinfo{journal}{Space Weather} \bibinfo{volume}{15}
  (\bibinfo{year}{2017}) \bibinfo{pages}{1039--1051}.
%Type = Article
\bibitem[{Stevenson et~al.(2022)Stevenson, Rodriguez-Fernandez, Minisci, and
  Camacho}]{stevenson2022}
\bibinfo{author}{E.~Stevenson}, \bibinfo{author}{V.~Rodriguez-Fernandez},
  \bibinfo{author}{E.~Minisci}, \bibinfo{author}{D.~Camacho},
\newblock \bibinfo{title}{A deep learning approach to solar radio flux
  forecasting},
\newblock \bibinfo{journal}{Acta Astronautica} \bibinfo{volume}{193}
  (\bibinfo{year}{2022}) \bibinfo{pages}{595--606}.
%Type = Article
\bibitem[{Camporeale(2019)}]{Camporeale2019}
\bibinfo{author}{E.~Camporeale},
\newblock \bibinfo{title}{{The Challenge of Machine Learning in Space Weather:
  Nowcasting and Forecasting}},
\newblock \bibinfo{journal}{Space Weather} \bibinfo{volume}{17}
  (\bibinfo{year}{2019}) \bibinfo{pages}{1166--1207}.
%Type = Article
\bibitem[{Doornbos et~al.(2008)Doornbos, Klinkrad, and
  Visser}]{DOORNBOS20081115}
\bibinfo{author}{E.~Doornbos}, \bibinfo{author}{H.~Klinkrad},
  \bibinfo{author}{P.~Visser},
\newblock \bibinfo{title}{Use of two-line element data for thermosphere neutral
  density model calibration},
\newblock \bibinfo{journal}{Advances in Space Research} \bibinfo{volume}{41}
  (\bibinfo{year}{2008}) \bibinfo{pages}{1115--1122}.
%Type = Article
\bibitem[{{Shi} et~al.(2015){Shi}, {Li}, {Li}, {Zhao}, and {Sang}}]{2015Shi}
\bibinfo{author}{C.~{Shi}}, \bibinfo{author}{W.~{Li}},
  \bibinfo{author}{M.~{Li}}, \bibinfo{author}{Q.~{Zhao}},
  \bibinfo{author}{J.~{Sang}},
\newblock \bibinfo{title}{{Calibrating the scale of the NRLMSISE00 model during
  solar maximum using the two line elements dataset}},
\newblock \bibinfo{journal}{Advances in Space Research} \bibinfo{volume}{56}
  (\bibinfo{year}{2015}) \bibinfo{pages}{1--9}.
%Type = Article
\bibitem[{Sang et~al.(2011)Sang, Smith, and Zhang}]{sang2011modification}
\bibinfo{author}{J.~Sang}, \bibinfo{author}{C.~Smith},
  \bibinfo{author}{K.~Zhang},
\newblock \bibinfo{title}{Modification of atmospheric mass density model
  coefficients using space tracking data--a simulation study for accurate
  debris orbit prediction},
\newblock \bibinfo{journal}{Advances in the Astronautical Sciences}
  \bibinfo{volume}{140} (\bibinfo{year}{2011}) \bibinfo{pages}{1479--1493}.
%Type = Article
\bibitem[{Chen et~al.(2019)Chen, Du, and Sang}]{Chen2019}
\bibinfo{author}{J.~Chen}, \bibinfo{author}{J.~Du}, \bibinfo{author}{J.~Sang},
\newblock \bibinfo{title}{{Improved orbit prediction of LEO objects with
  calibrated atmospheric mass density model}},
\newblock \bibinfo{journal}{Journal of Spatial Science} \bibinfo{volume}{64}
  (\bibinfo{year}{2019}) \bibinfo{pages}{97--110}.
%Type = Article
\bibitem[{Elvidge et~al.(2016)Elvidge, Godinez, and Angling}]{Elvidge2016}
\bibinfo{author}{S.~Elvidge}, \bibinfo{author}{H.~C. Godinez},
  \bibinfo{author}{M.~J. Angling},
\newblock \bibinfo{title}{{Improved forecasting of thermospheric densities
  using multi-model ensembles}},
\newblock \bibinfo{journal}{Geoscientific Model Development}
  \bibinfo{volume}{9} (\bibinfo{year}{2016}) \bibinfo{pages}{2279--2292}.
%Type = Article
\bibitem[{P{\'{e}}rez and Bevilacqua(2015)}]{Perez2015}
\bibinfo{author}{D.~P{\'{e}}rez}, \bibinfo{author}{R.~Bevilacqua},
\newblock \bibinfo{title}{{Neural Network based calibration of atmospheric
  density models}},
\newblock \bibinfo{journal}{Acta Astronautica} \bibinfo{volume}{110}
  (\bibinfo{year}{2015}) \bibinfo{pages}{58--76}.
%Type = Article
\bibitem[{Chen et~al.(2014)Chen, Liu, and Hanada}]{Chen2014}
\bibinfo{author}{H.~Chen}, \bibinfo{author}{H.~Liu},
  \bibinfo{author}{T.~Hanada},
\newblock \bibinfo{title}{Storm-time atmospheric density modeling using neural
  networks and its application in orbit propagation},
\newblock \bibinfo{journal}{Advances in Space Research} \bibinfo{volume}{53}
  (\bibinfo{year}{2014}) \bibinfo{pages}{558--567}.
%Type = Article
\bibitem[{Zhang et~al.(2021)Zhang, Yu, Chen, and Sang}]{Zhang2021}
\bibinfo{author}{Y.~Zhang}, \bibinfo{author}{J.~Yu}, \bibinfo{author}{J.~Chen},
  \bibinfo{author}{J.~Sang},
\newblock \bibinfo{title}{An empirical atmospheric density calibration model
  based on long short-term memory neural network},
\newblock \bibinfo{journal}{Atmosphere} \bibinfo{volume}{12}
  (\bibinfo{year}{2021}).
%Type = Article
\bibitem[{Gao et~al.(2020)Gao, Peng, and Bai}]{GAO2020273}
\bibinfo{author}{T.~Gao}, \bibinfo{author}{H.~Peng}, \bibinfo{author}{X.~Bai},
\newblock \bibinfo{title}{Calibration of atmospheric density model based on
  gaussian processes},
\newblock \bibinfo{journal}{Acta Astronautica} \bibinfo{volume}{168}
  (\bibinfo{year}{2020}) \bibinfo{pages}{273--281}.
%Type = Inproceedings
\bibitem[{Mehta and Linares(2019)}]{Mehta2018}
\bibinfo{author}{P.~M. Mehta}, \bibinfo{author}{R.~Linares},
\newblock \bibinfo{title}{{Data-Driven Framework for Real-time Thermospheric
  Density Estimation}},
\newblock in: \bibinfo{booktitle}{Advances in the Astronautical Sciences
  Conference 2019}, volume \bibinfo{volume}{167}, \bibinfo{address}{Portland,
  ME}, \bibinfo{year}{2019}, pp. \bibinfo{pages}{191--207}.
%Type = Article
\bibitem[{Mehta and Linares(2018)}]{Mehta_2018}
\bibinfo{author}{P.~M. Mehta}, \bibinfo{author}{R.~Linares},
\newblock \bibinfo{title}{A new transformative framework for data assimilation
  and calibration of physical ionosphere-thermosphere models},
\newblock \bibinfo{journal}{Space Weather} \bibinfo{volume}{16}
  (\bibinfo{year}{2018}) \bibinfo{pages}{1086--1100}.
%Type = Article
\bibitem[{Gondelach and Linares(2020)}]{Gondelach2020}
\bibinfo{author}{D.~J. Gondelach}, \bibinfo{author}{R.~Linares},
\newblock \bibinfo{title}{Real-time thermospheric density estimation via
  two-line element data assimilation},
\newblock \bibinfo{journal}{Space Weather} \bibinfo{volume}{18}
  (\bibinfo{year}{2020}) \bibinfo{pages}{20}.
%Type = Article
\bibitem[{Gondelach and Linares(2021)}]{Gondelach2021}
\bibinfo{author}{D.~J. Gondelach}, \bibinfo{author}{R.~Linares},
\newblock \bibinfo{title}{{Real-Time Thermospheric Density Estimation Via Radar
  and GPS Tracking Data Assimilation}},
\newblock \bibinfo{journal}{Space Weather} \bibinfo{volume}{19}
  (\bibinfo{year}{2021}) \bibinfo{pages}{18}.
%Type = Inproceedings
\bibitem[{Turner et~al.(2020)Turner, Zhang, Gondelach, and
  Linares}]{Turner2020}
\bibinfo{author}{H.~Turner}, \bibinfo{author}{M.~Zhang},
  \bibinfo{author}{D.~Gondelach}, \bibinfo{author}{R.~Linares},
\newblock \bibinfo{title}{Machine learning algorithms for improved
  thermospheric density modeling},
\newblock in: \bibinfo{booktitle}{Dynamic Data Driven Applications Systems},
  \bibinfo{publisher}{Springer International Publishing},
  \bibinfo{address}{Cham}, \bibinfo{year}{2020}, pp. \bibinfo{pages}{143--151}.
%Type = Inproceedings
\bibitem[{Nateghi et~al.(2021)Nateghi, Manzi, and Vasile}]{Nateghi2021}
\bibinfo{author}{V.~Nateghi}, \bibinfo{author}{M.~Manzi},
  \bibinfo{author}{M.~Vasile},
\newblock \bibinfo{title}{{Autoencoder-Based Thermospheric Density Estimation
  Using GPS Tracking Data}},
\newblock in: \bibinfo{booktitle}{72nd International Astronautical Congress},
  \bibinfo{publisher}{IAF}, \bibinfo{address}{Dubai, United Arab Emirates},
  \bibinfo{year}{2021}, p.~\bibinfo{pages}{10}.
%Type = Article
\bibitem[{Mehta et~al.(2017)Mehta, Walker, Sutton, and Godinez}]{Mehta2017}
\bibinfo{author}{P.~M. Mehta}, \bibinfo{author}{A.~C. Walker},
  \bibinfo{author}{E.~K. Sutton}, \bibinfo{author}{H.~C. Godinez},
\newblock \bibinfo{title}{{New density estimates derived using accelerometers
  on board the CHAMP and GRACE satellites}},
\newblock \bibinfo{journal}{Space Weather} \bibinfo{volume}{15}
  (\bibinfo{year}{2017}) \bibinfo{pages}{558--576}.
%Type = Article
\bibitem[{George and McLaughlin(2021)}]{George2021}
\bibinfo{author}{T.~R. George}, \bibinfo{author}{C.~A. McLaughlin},
\newblock \bibinfo{title}{{The Use of Long Short-Term Memory Artificial Neural
  Networks for the Global Prediction of Atmospheric Density}},
\newblock \bibinfo{journal}{Advances in the Astronautical Sciences}
  \bibinfo{volume}{175} (\bibinfo{year}{2021}) \bibinfo{pages}{1815--1832}.
%Type = Inproceedings
\bibitem[{Bonasera et~al.(2021)Bonasera, P{\'{e}}rez-hern{\'{a}}ndez, Brown,
  Acciarini, Benson, Sutton, and Bridges}]{Bonasera2021}
\bibinfo{author}{S.~Bonasera}, \bibinfo{author}{J.~A.
  P{\'{e}}rez-hern{\'{a}}ndez}, \bibinfo{author}{E.~Brown},
  \bibinfo{author}{G.~Acciarini}, \bibinfo{author}{B.~Benson},
  \bibinfo{author}{E.~Sutton}, \bibinfo{author}{C.~Bridges},
\newblock \bibinfo{title}{{Dropout and Ensemble Networks for Thermospheric
  Density Uncertainty Estimation}},
\newblock in: \bibinfo{booktitle}{Bayesian Deep Learning Workshop},
  \bibinfo{number}{Dec.}, \bibinfo{publisher}{NeurIPS}, \bibinfo{year}{2021},
  pp. \bibinfo{pages}{1--7}.
%Type = Inproceedings
\bibitem[{Young et~al.(2018)Young, Abdou, and Bener}]{Steven2018}
\bibinfo{author}{S.~Young}, \bibinfo{author}{T.~Abdou},
  \bibinfo{author}{A.~Bener},
\newblock \bibinfo{title}{Deep super learner: A deep ensemble for
  classification problems},
\newblock in: \bibinfo{booktitle}{Advances in Artificial Intelligence},
  \bibinfo{publisher}{Springer International Publishing},
  \bibinfo{address}{Cham}, \bibinfo{year}{2018}, pp. \bibinfo{pages}{84--95}.
%Type = Inproceedings
\bibitem[{Benson et~al.(2021)Benson, Brown, Bonasera, Acciarini,
  P{\'{e}}rez-hern{\'{a}}ndez, Sutton, Jah, Bridges, Jin, and {G{\"{u}}nes
  Baydin}}]{Benson2021}
\bibinfo{author}{B.~Benson}, \bibinfo{author}{E.~Brown},
  \bibinfo{author}{S.~Bonasera}, \bibinfo{author}{G.~Acciarini},
  \bibinfo{author}{J.~A. P{\'{e}}rez-hern{\'{a}}ndez},
  \bibinfo{author}{E.~Sutton}, \bibinfo{author}{M.~K. Jah},
  \bibinfo{author}{C.~Bridges}, \bibinfo{author}{M.~Jin},
  \bibinfo{author}{A.~{G{\"{u}}nes Baydin}},
\newblock \bibinfo{title}{{Simultaneous Multivariate Forecast of Space Weather
  Indices using Deep Neural Network Ensembles}},
\newblock in: \bibinfo{booktitle}{Fourth Workshop on Machine Learning and the
  Physical Sciences}, \bibinfo{publisher}{NeurIPS}, \bibinfo{year}{2021}, pp.
  \bibinfo{pages}{1--6}.
%Type = Article
\bibitem[{Tobiska et~al.(2021)Tobiska, Bowman, Bouwer, Cruz, Wahl, Pilinski,
  Mehta, and Licata}]{Tobiska2021}
\bibinfo{author}{W.~K. Tobiska}, \bibinfo{author}{B.~R. Bowman},
  \bibinfo{author}{S.~D. Bouwer}, \bibinfo{author}{A.~Cruz},
  \bibinfo{author}{K.~Wahl}, \bibinfo{author}{M.~D. Pilinski},
  \bibinfo{author}{P.~M. Mehta}, \bibinfo{author}{R.~J. Licata},
\newblock \bibinfo{title}{{The SET HASDM Density Database}},
\newblock \bibinfo{journal}{Space Weather} \bibinfo{volume}{19}
  (\bibinfo{year}{2021}) \bibinfo{pages}{1--4}.
%Type = Article
\bibitem[{Licata et~al.(2022)Licata, Mehta, Tobiska, and
  Huzurbazar}]{Licata2022}
\bibinfo{author}{R.~J. Licata}, \bibinfo{author}{P.~M. Mehta},
  \bibinfo{author}{W.~K. Tobiska}, \bibinfo{author}{S.~Huzurbazar},
\newblock \bibinfo{title}{{Machine‐Learned HASDM Thermospheric Mass Density
  Model With Uncertainty Quantification}},
\newblock \bibinfo{journal}{Space Weather} \bibinfo{volume}{20}
  (\bibinfo{year}{2022}) \bibinfo{pages}{1--18}.
%Type = Article
\bibitem[{Licata and Mehta(2022)}]{Licata2022a}
\bibinfo{author}{R.~J. Licata}, \bibinfo{author}{P.~M. Mehta},
\newblock \bibinfo{title}{{Uncertainty Quantification Techniques for Space
  Weather Modeling: Thermospheric Density Application}},
\newblock \bibinfo{journal}{Scientific Reports} \bibinfo{volume}{12}
  (\bibinfo{year}{2022}) \bibinfo{pages}{1--17}.
%Type = Incollection
\bibitem[{Goan and Fookes(2020)}]{Goan_2020}
\bibinfo{author}{E.~Goan}, \bibinfo{author}{C.~Fookes},
\newblock \bibinfo{title}{Bayesian neural networks: An introduction and
  survey},
\newblock in: \bibinfo{booktitle}{Case Studies in Applied Bayesian Data
  Science}, \bibinfo{publisher}{Springer International Publishing},
  \bibinfo{address}{Cham}, \bibinfo{year}{2020}, pp. \bibinfo{pages}{45--87}.
%Type = Article
\bibitem[{Hall(2021)}]{Hall2021}
\bibinfo{author}{D.~T. Hall},
\newblock \bibinfo{title}{Expected collision rates for tracked satellites},
\newblock \bibinfo{journal}{Journal of Spacecraft and Rockets}
  \bibinfo{volume}{58} (\bibinfo{year}{2021}) \bibinfo{pages}{715--728}.
%Type = Phdthesis
\bibitem[{Chipade(2021)}]{Chipade2021}
\bibinfo{author}{R.~Chipade}, \bibinfo{title}{Contributions to the Statistical
  Orbit Determination of Satellites}, Ph.D. thesis, Savitribai Phule Pune
  University, \bibinfo{address}{Pune, India}, \bibinfo{year}{2021}.
%Type = Article
\bibitem[{Guthrie et~al.(2022)Guthrie, Kim, Urrutxua, and Hare}]{GUTHRIE2022}
\bibinfo{author}{B.~Guthrie}, \bibinfo{author}{M.~Kim},
  \bibinfo{author}{H.~Urrutxua}, \bibinfo{author}{J.~Hare},
\newblock \bibinfo{title}{Image-based attitude determination of co-orbiting
  satellites using deep learning technologies},
\newblock \bibinfo{journal}{Aerospace Science and Technology}
  \bibinfo{volume}{120} (\bibinfo{year}{2022}) \bibinfo{pages}{14}.
%Type = Article
\bibitem[{Sharma and D’Amico(2020)}]{Sharma}
\bibinfo{author}{S.~Sharma}, \bibinfo{author}{S.~D’Amico},
\newblock \bibinfo{title}{Neural network-based pose estimation for
  noncooperative spacecraft rendezvous},
\newblock \bibinfo{journal}{Institute of Electric and Electronics Engineers
  (IEEE) Transactions on Aerospace and Electronic Systems} \bibinfo{volume}{56}
  (\bibinfo{year}{2020}) \bibinfo{pages}{4638--4658}.
%Type = Inproceedings
\bibitem[{Chen et~al.(2019)Chen, Cao, Parra, and Chin}]{chen2019satellite}
\bibinfo{author}{B.~Chen}, \bibinfo{author}{J.~Cao},
  \bibinfo{author}{A.~Parra}, \bibinfo{author}{T.~Chin},
\newblock \bibinfo{title}{Satellite pose estimation with deep landmark
  regression and nonlinear pose refinement},
\newblock in: \bibinfo{booktitle}{2019 IEEE/CVF International Conference on
  Computer Vision Workshop (ICCVW)}, \bibinfo{publisher}{IEEE Computer
  Society}, \bibinfo{address}{Los Alamitos, CA, USA}, \bibinfo{year}{2019}, pp.
  \bibinfo{pages}{2816--2824}.
%Type = Article
\bibitem[{Linares et~al.(2020)Linares, Furfaro, and Reddy}]{Furfaro2019}
\bibinfo{author}{R.~Linares}, \bibinfo{author}{R.~Furfaro},
  \bibinfo{author}{V.~Reddy},
\newblock \bibinfo{title}{Space objects classification via light-curve
  measurements using deep convolutional neural networks},
\newblock \bibinfo{journal}{The Journal of the Astronautical Sciences}
  \bibinfo{volume}{67} (\bibinfo{year}{2020}) \bibinfo{pages}{1063--1091}.
%Type = Inproceedings
\bibitem[{{Furfaro} et~al.(2019){Furfaro}, {Campbell}, {Linares}, and
  {Reddy}}]{Furfaro2019a}
\bibinfo{author}{R.~{Furfaro}}, \bibinfo{author}{T.~{Campbell}},
  \bibinfo{author}{R.~{Linares}}, \bibinfo{author}{V.~{Reddy}},
\newblock \bibinfo{title}{{Space Debris Identification and Characterization via
  Deep Meta-Learning}},
\newblock in: \bibinfo{booktitle}{First International Orbital Debris
  Conference}, volume \bibinfo{volume}{2109} of \textit{\bibinfo{series}{LPI
  Contributions}}, \bibinfo{address}{Sugar Land,TX}, \bibinfo{year}{2019},
  p.~\bibinfo{pages}{9}.

\end{thebibliography}

%% else use the following coding to input the bibitems directly in the
%% TeX file.

% \begin{thebibliography}{00}

% %% \bibitem{label}
% %% Text of bibliographic item

% \bibitem{}

% \end{thebibliography}
\end{document}